\documentclass[12pt,preprint]{aastex}

\begin{document}

\title{A Suborbital Payload for Soft X-ray Spectroscopy of Extended Sources}

\author{P.H.H. Oakley$^1$, R.L. McEntaffer$^2$, \& W. Cash$^1$}

\affil{$^1$Center for Astrophysics and Space Astronomy, University of Colorado, Boulder, CO 80309}
\affil{$^2$Department of Physics and Astronomy, University of Iowa, Iowa City, IA 52242}
\email{Phil.Oakley@Colorado.edu}

\begin{abstract}
We present a suborbital rocket payload capable of performing soft X-ray spectroscopy on extended sources. The payload can reach resolutions of $\sim100$ ($\lambda / \Delta \lambda$) over sources as large as $3.25^{\circ}$ in diameter in the 17-107 \AA\ bandpass. This permits analysis of the overall energy balance of nearby supernova remnants and the detailed nature of the diffuse soft X-ray background. The main components of the instrument are: wire grid collimators, off-plane grating arrays and gaseous electron multiplier detectors. This payload is adaptable to longer duration orbital rockets given its comparatively simple pointing and telemetry requirements and an abundance of potential science targets.

\keywords{suborbital rockets \and X-ray spectroscopy \and Gaseous electron multipliers \and off-plane gratings \and X-ray detectors \and grazing incidence optics}

\end{abstract}

\section{Introduction}
\label{intro}
The ROSAT all sky survey imaged a wealth of extended soft X-ray emission and highlighted the need for a high resolution extended spectroscopic instrument (\cite{Snowden1995} and \cite{Snowden1997}).  Potential science includes probing the composition and evolution of supernova remnants (\cite{Koyama1995}, \cite{Levenson2005}, \cite{Flanagan2004}, \cite{Reynolds1999}), studying the specifics of charge exchange of the solar wind with interstellar neutrals (\cite{Lisse1996}, \cite{Cravens1997}, \cite{Cravens2001}), determining the nature of emission from the galactic halo, and obtaining further diagnostics of the soft X-ray background (\cite{Sanders1998}, \cite{Sanders2001}, \cite{McCammon2002}, \cite{Cox2005}).

The current generation of X-ray observatories (Chandra, XMM-Newton, Suzaku, etc.) are capable of producing impressive images of point sources and moderately extended sources. However the data gathered from these images have poor spectral information due to the limited energy resolution of the CCD. For Chandra this is approximately 100 eV (\cite{Garmire2003} and \cite{Chandra2009}), giving poor resolution (equivalent to $\sim<10 \; \lambda / \Delta \lambda)$ for photons $<1$ keV. Smaller objects ($<1^\prime$) can be analyzed via the onboard gratings, as shown successfully by \cite{Flanagan2004}. However, the resolution is still limited by the intrinsic angular size of the object.  For many important X-ray sources such as galactic supernova remnants, galactic halo emission, the LHB, and the local soft X-ray background, this technique is impractical. Thus, a new spectrometer design is needed for high spectral resolution of larger sources. 

A powerful instrument was built to suit these needs with the Cygnus X-ray Emission Spectroscopic Survey (CyXESS), flown in 2006 \cite{McEntaffer2006}. This instrument successfully observed the Cygnus Loop SNR \cite{McEntaffer2008}. The Extended X-ray Off-plane Spectrometer (EXOS) sounding rocket payload was an upgrade of the existing CyXESS payload, modified to provide higher sensitivity and lower noise observations \cite{Oakley2009}, \cite{Oakley2010}. This payload was launched successfully in late 2009 to re-observe the Cygnus Loop (\cite{Blair1999}, \cite{Levenson1998}, \cite{Levenson2005}, \cite{Tsunemi2009}, \cite{Katsuda2008}). Unfortunately, the EXOS payload was damaged upon landing, necessitating a rebuild before relaunch. This rebuild, known as the CODEX payload, has been designed to further improve upon the overall throughput. Our description below describes all three instrument versions.  The following sections provide an overview of the instrumental design (Section 2), its performance in the lab and field (Sections 3 and 4) and our plans for future flights (Section 5). 

\section{Instrument Description}
\label{sec:1}
The main optical components of this spectrometer design are a wire grid collimator, an off-plane reflection grating array, and Gaseous Electron Multiplier (GEM) detectors.  The payload has two identical modules, each containing a collimator, grating and GEM detector.  Each component will be discussed in detail below. Table \ref{table:exos_parameters} shows a list of relevant EXOS parameters.

\subsection{Wire-Grid Collimator}
Wire grids placed along the optical axis can be used to filter out converging and diverging light, allowing only collimated light to pass through the system. This system is known as a wire-grid collimator and has been widely and effectively used in X-ray astronomy (e.g. \cite{Gunderson2000}). We use a similar structure to manipulate what light passes through our system. Instead of allowing only collimated light through the system, we allow only light travelling towards the desired position on the focal plane to pass unimpeded. The image from each of the slits converge at the focal plane, creating a single line. All other photons are vignetted by the wire bars. 

\begin{figure}
  \includegraphics[width=0.8\textwidth]{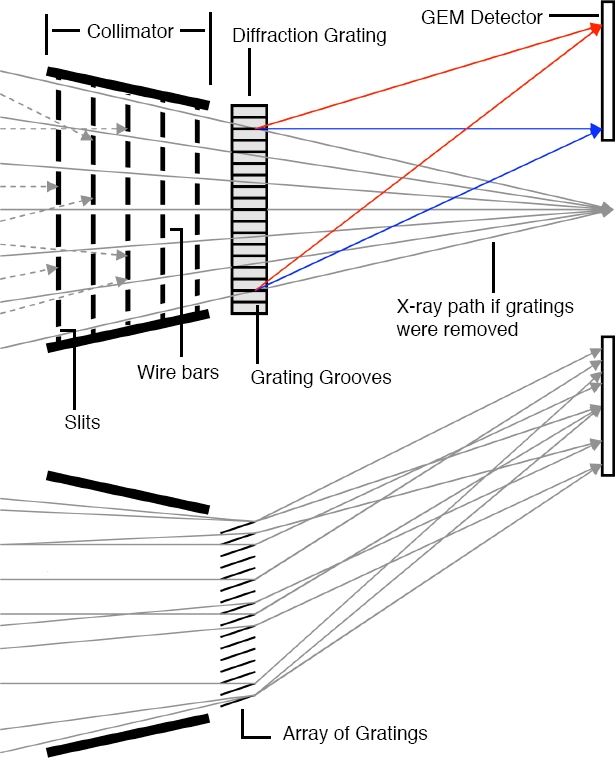}
\caption{Top - Photons travelling towards the desired focal line are allowed to pass through the slits. Photons not travelling towards this focus (shown with dashed lines) are vignetted by the wire bars. The red and blue lines show the path the gratings disperse. The angle formed by the collimator defines the FOV of the system. Bottom - View along the orthogonal axis as the X-rays travel down one slit. Along this axis the photons within the FOV are not collimated, resulting in a thin line at the focal plane rather than a point. This system provides no angular resolution on the astronomical target. Drawing is not to scale.}
\label{fig:1}      
\end{figure}

A schematic of our ``converging collimator" or ``slit overlapper" is shown in Figure \ref{fig:1}. Wire grids are placed along the optical axis with successively smaller slit width between the wire bars (which also decrease in width). Light travels from the entrance slits through the system, encountering slits on each plate that vignette any rays not travelling towards the desired focus. These plates are arranged from the aperture down to approximately one meter of depth along the optical axis. The location of each wire-grid plate is determined by the raytace of the system. This raytrace places each plate at an optical depth that prevents light from entering a neighboring slit on the next plate. There are 24 total plates per module.

Each slit in the collimator sees a different portion of the sky, however, when added together, the overall field of view (FOV) of the system is $3.25^{\circ}$ x $3.25^{\circ}$. These plates sculpt a converging beam in only one dimension, creating a focal line rather than a spot. This system is designed to obtain spectra of large extended sources and provides no angular resolution.

The focal length of the system is three meters, but the first meter of sculpting creates a beam to a full width half max (FWHM) of 1.6 mm with a scatter at the level of $\sim1\%$. Theoretically this beam could be sculpted with no scatter. The photon/wire encounters occur at roughly normal incidence, causing undesired light to be absorbed and removed from the beam rather than scattered into the system. However, optimal placement of each plate places them in difficult or impossible proximity to other grids, thus producing grid spacings of higher precision than possible with current techniques.

\begin{figure}
  \includegraphics[width=\textwidth]{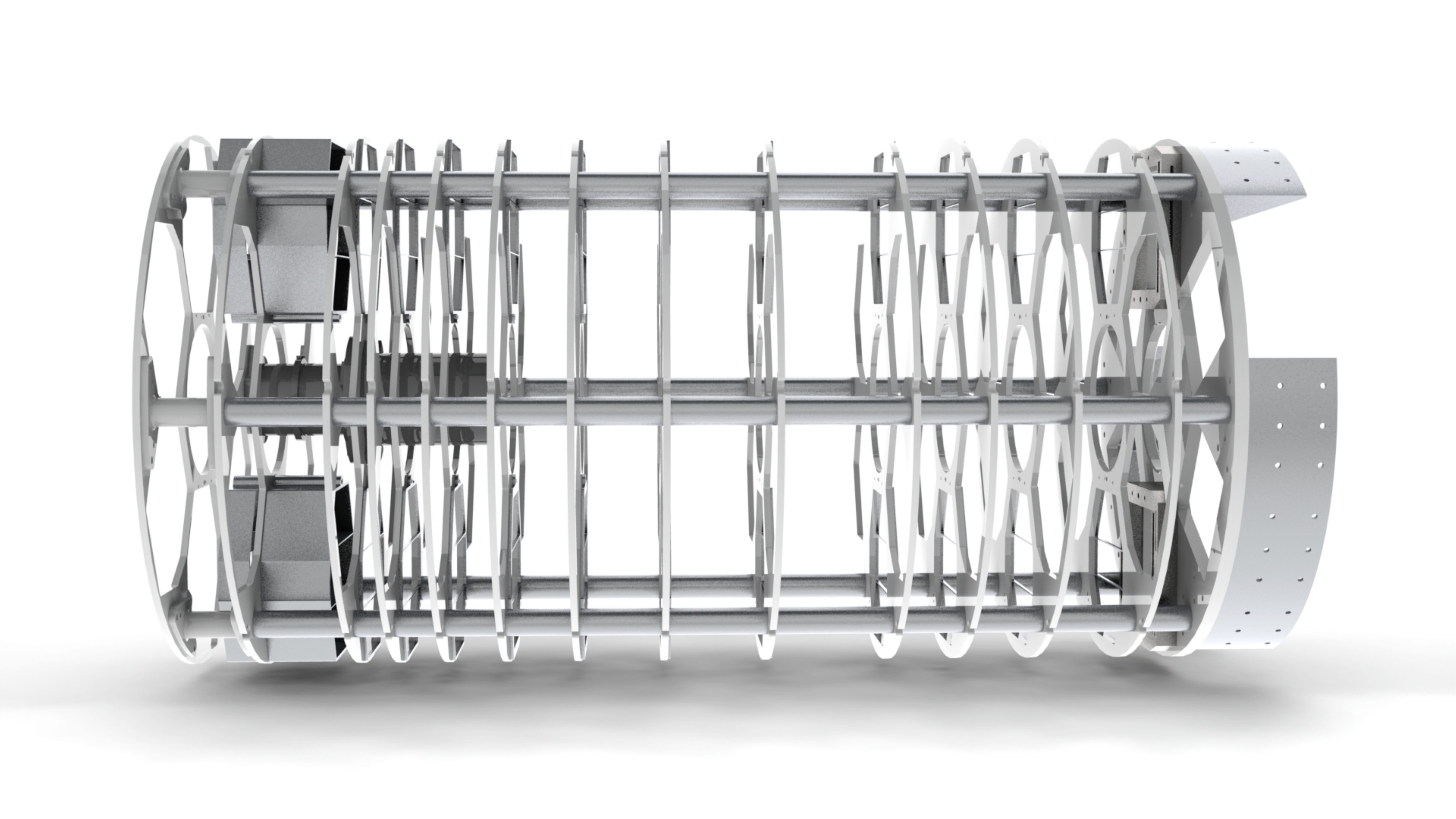}
\caption{A SolidWorks 3D rendering of the super-structure to which the wire-grid plates are mounted. The two leftmost plates have multiple grids mounted on them, while the remaining plates support only one grid apiece.}
\label{fig:2}      
\end{figure}

This grid system has several practical benefits over the use of traditional mirror based designs. These grazing incidence optics are expensive and difficult to produce. To achieve enough collecting area they must be made thin and nested into arrays. In addition to the cost, this requires complicated mounting structures to achieve and maintain alignment through the vibrations of launch. A wire-grid collimator is inexpensive, easier to align, and can be mounted simply and securely.  The drawbacks are that the focus is in only one dimension and not as fine as those achieved by reflective optics. Figure \ref{fig:2} shows an engineering model rendering of the whole structure.

\begin{figure}
  \includegraphics[width=0.5\textwidth]{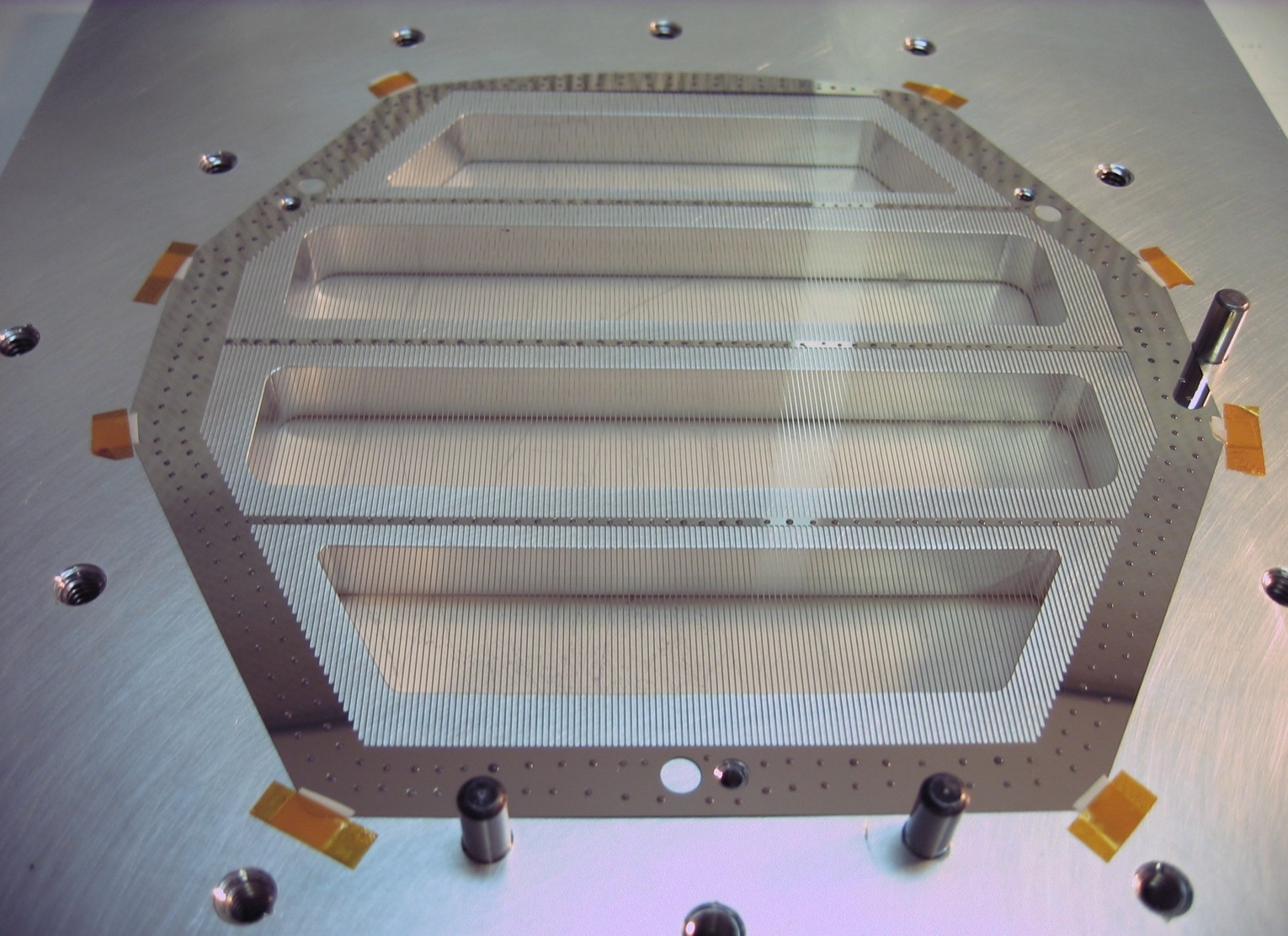}
  \includegraphics[width=0.5\textwidth]{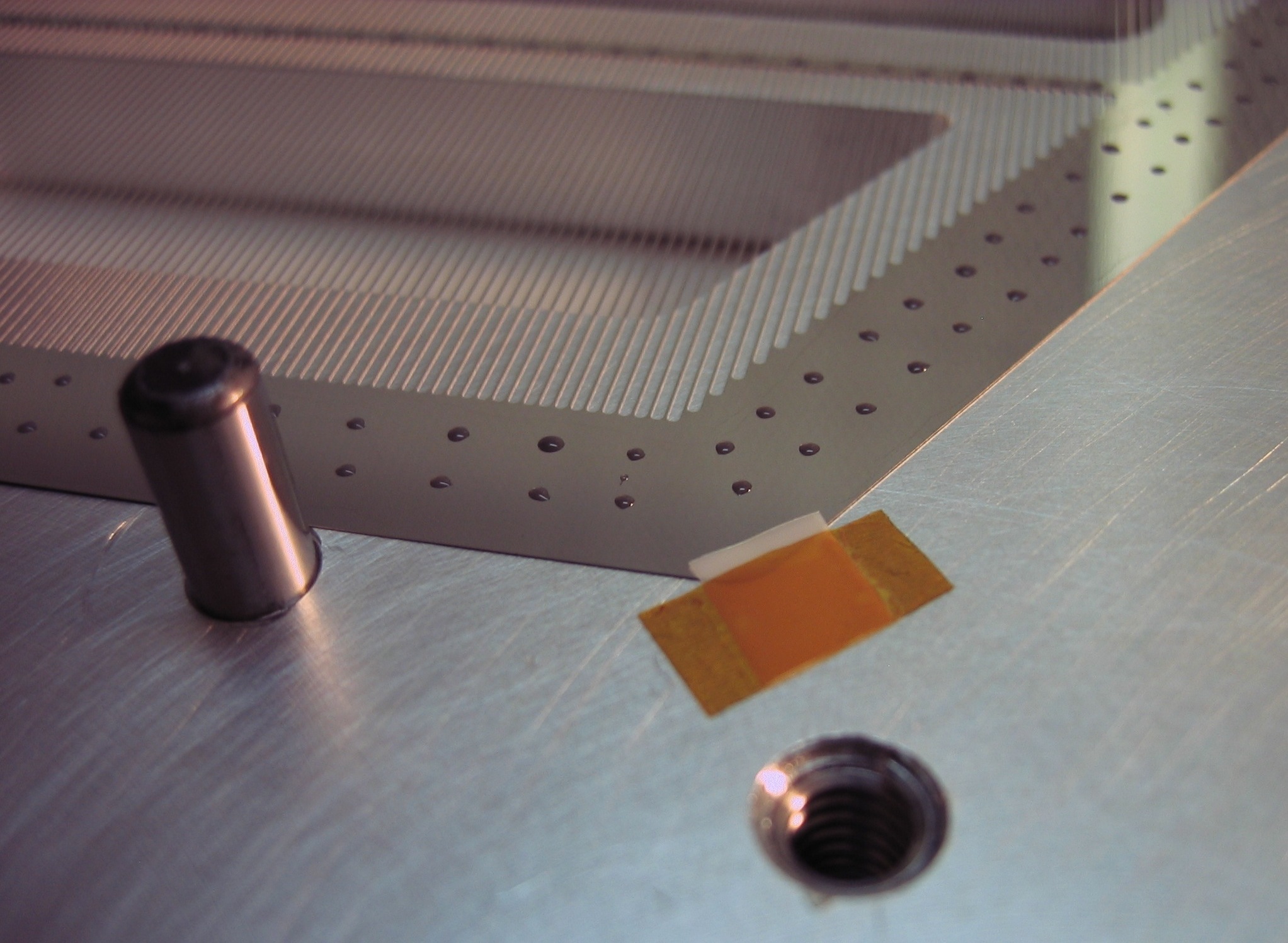}
\caption{Left - A collimator plate on its bonding fixture. Right - Close-up view of one of the plates. The wire bars (and slits) run approximately vertically in these photographs. Each module has 24 of these plates distributed at intervals over the 1m length of the structure.}
\label{fig:3}      
\end{figure}

The wire grid plates are created via electroforming nickel and then mounted on machined aluminum frames for support. The initial opening size between wire bars is set at 725 microns while the final plate has a slit size of 500 microns. The wire bars themselves decrease from a width of 166 $\mu$m to 114 $\mu$m at the bottom. These wire bars are capable of withstanding $>1$ pound of force each before yielding. There are 185 slits per plate. The plates are 6.745 by 6.575 inches and require three cross braces for structural support. These cross braces reduce the unsupported length of the wire bars, thus decreasing the maximum deflection expected during launch. This helps avoid fatigue on the wire bars which would degrade optical performance. A photograph of one of these plates prior to being bonded to its aluminum support frame is shown in Figure \ref{fig:3}.

\begin{figure}
  \includegraphics[width=0.75\textwidth]{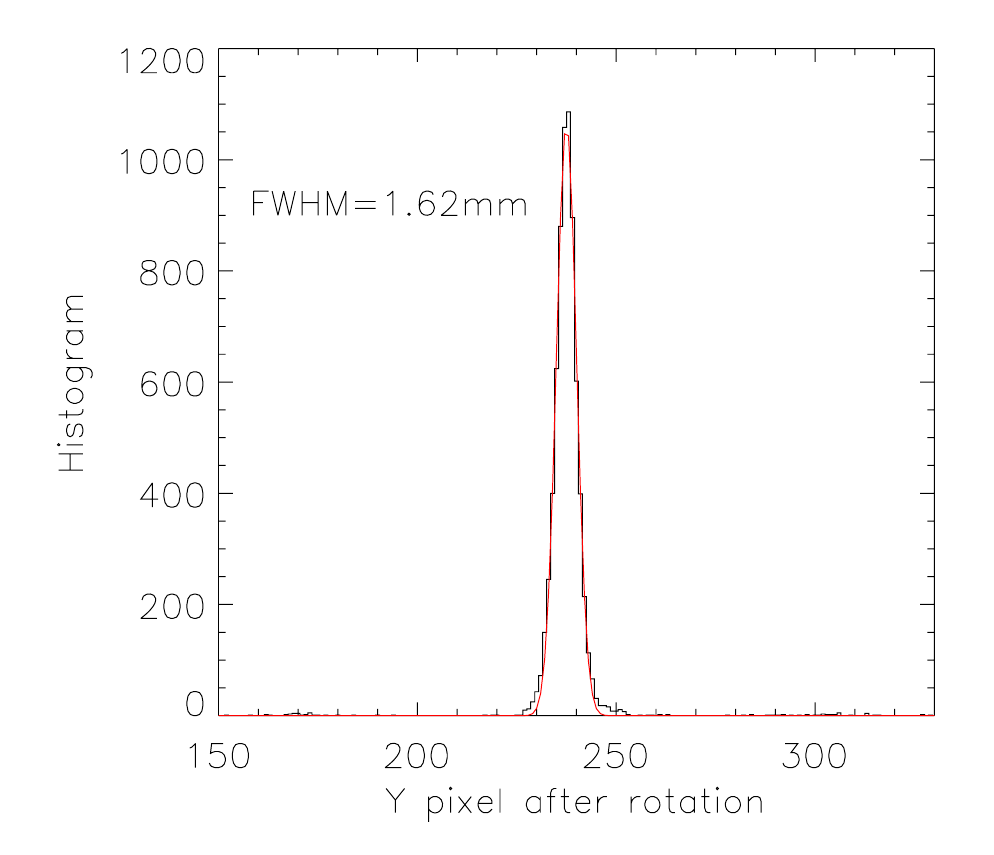}
\caption{Results of collimator calibration data. The red line indicates the best fit Gaussian curve. The observed FWHM matches the raytrace at $\sim1.6$mm. The amount of scatter is also as expected at $\sim1\%$.}
\label{fig:4}      
\end{figure}

The telescope aperture is defined by the extent of our target, the Cygnus Loop Supernova Remnant. In order to maximize the number of these wire grid collimator modules that will fit within the limited payload envelope ($22''$ in diameter), the plates are shaped octagonally.  The bonding process is designed to precisely attach the aluminum frame without covering any effective area or inducing stress into the assembly that would cause the wire bars to warp. The plate is positioned against three precision pins. Low outgassing epoxy (2216 Scotch Weld) is mixed with precision sized glass-beads ($0.0025^{''}$) to create an equal depth bond line of epoxy around the entire plate. The amount of glass beads was set at $5\%$ (by volume) which had been shown to maximize shear strength through a series of stress tests. The frame was then placed on top of the epoxy and lightly clamped until the epoxy cured.  This assembly was then mounted in the collimator super-structure using a set of 5 optical lasers. Three of these lasers were aligned to shine up the central slit of the system, while the other two were set to each side of the center line and angled to hit the line defined by the three lasers on the focal plane. This ensured the plates are placed to prevent any relative rotation along the optical axis and lateral shifting between plates. Figure \ref{fig:4} shows the resulting point spread function of the wire-grid collimator.

\subsection{Off-Plane Gratings}
\begin{figure}
  \includegraphics[width=0.8\textwidth]{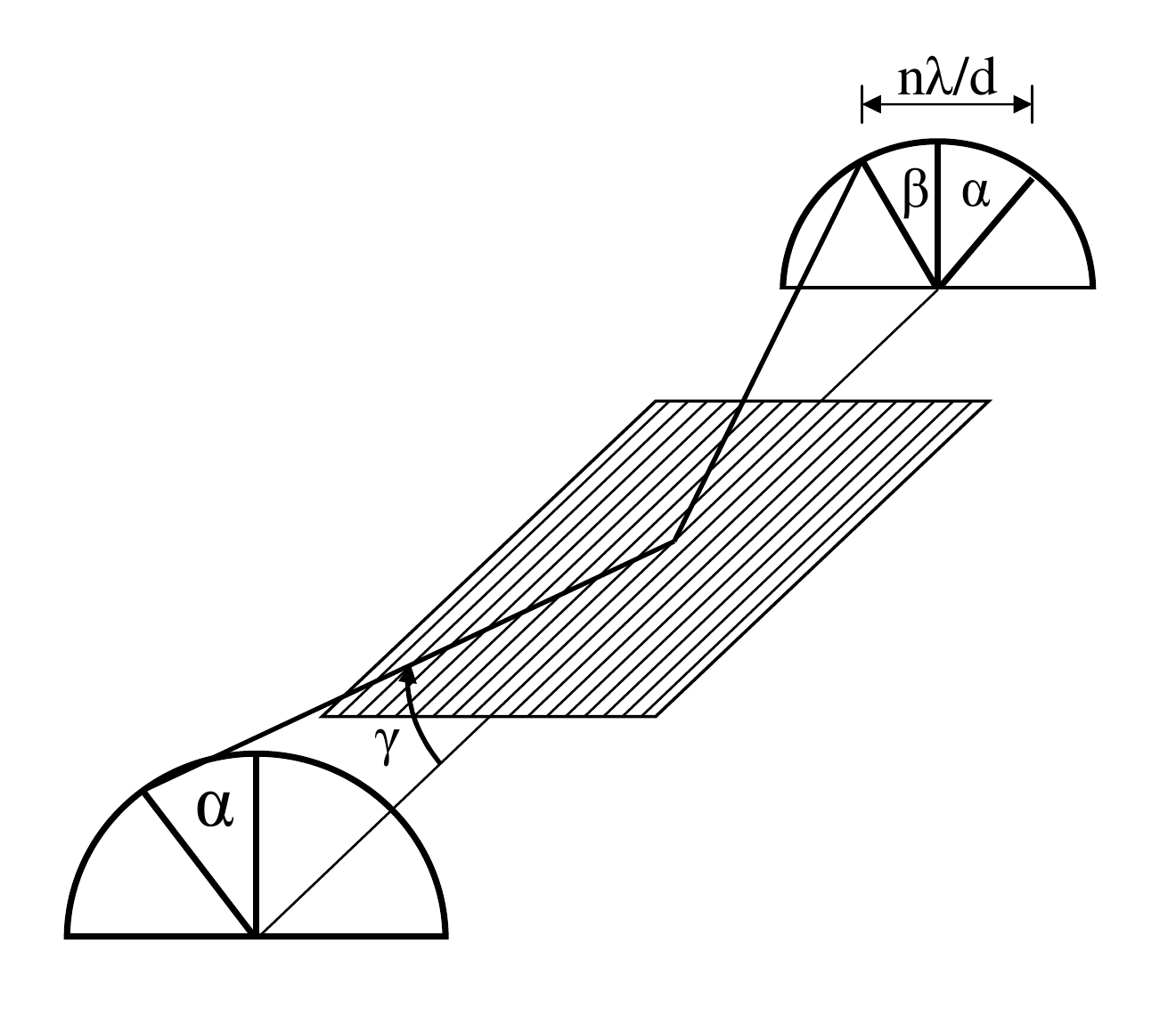}
\caption{Off plane grating geometry. The grooves are represented by the lines, while gamma represents the graze angle (4.4 degrees). Alpha and beta represent the incoming and outgoing diffracted angle. With this geometry we achieve an arc of diffraction whose radius is determined by gamma and the throw length (distance from reflection to focal plane).}
\label{fig:5}      
\end{figure}

After approximately a meter of travel within the collimator structure, the beam is still substantially large (104 x 104 mm). This makes it impractical to diffract the beam with a single grating without resorting to high graze angles (and thus low efficiency) or impractically large gratings. Thus we need an array of thin gratings to capture and properly diffract the entire beam with minimal loss. These gratings are held in tension with 5 lb of force to maintain flatness within one part in 2000 along their length and to prevent gratings from hitting each other during launch vibrations. The gratings were designed in the off-plane geometry (\cite{Cash1982}) where light approaches the gratings quasi-parallel to the grooves (Figure \ref{fig:5}).  This geometry was highly desirable for many reasons. With in-plane geometry one experiences a drop in efficiency due to groove shadowing that is avoided by choosing the high efficiency off-plane mount (\cite{Werner1977}). An in-plane setup could also diffract light into orders that intersect the next grating within the array, thus losing these photons. The off-plane mount disperses light conically at the shallow graze angle ($4.4^{\circ}$ in our case) allowing capture of all diffracted orders (see \cite{Osterman2004} for an example). Additionally, optical errors in fabrication and assembly create blurs that are almost entirely in the in-plane direction.  Since the off-plane disperses perpendicular to this direction, there is a significant easing of fabrication tolerances, and the packing geometries can be substantially better \cite{Cash1991}. Off-plane gratings have potential for higher resolution work and are currently being studied for the International X-ray Observatory (IXO) (\cite{McEntaffer2008}, \cite{McEntaffer2009}, \cite{Casement2010}, \cite{McEntaffer2010}).

\begin{figure}
  \includegraphics[width=0.9\textwidth]{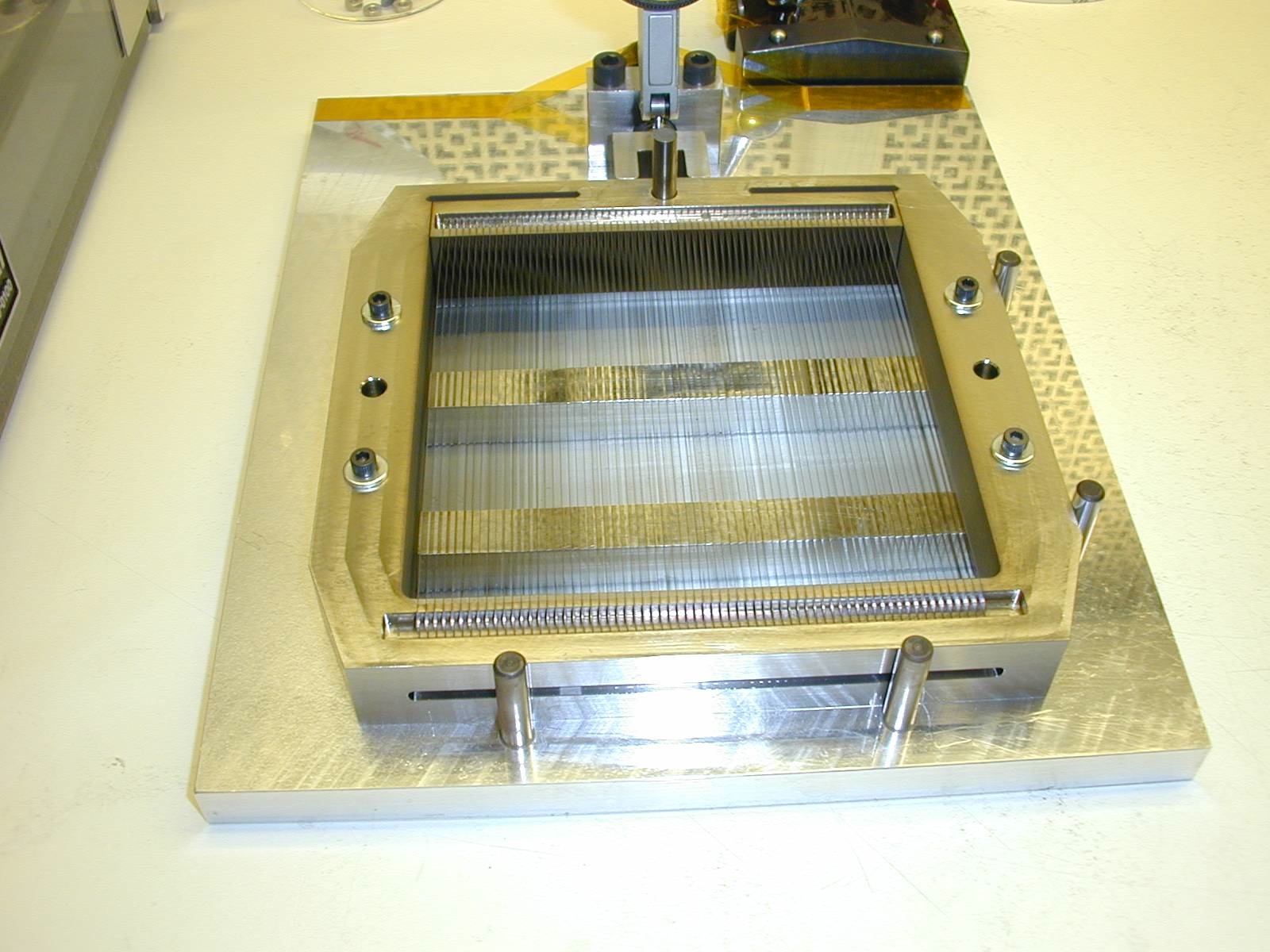}
\caption{Grating array prior to installation in the payload. The array consists of 67 gratings held in tension to ensure flatness. The grating substrates are electroformed nickel and are secured on a titanium flexure mount prior to loading.}
\label{fig:6}      
\end{figure}

The grating array contains 67 individual gratings (Figure \ref{fig:6}) per module. To minimize the vignetted light due to rays striking the edge of the grating, we used electroformed nickel for our substrate material. These substrates are formed to a thinness of $0.005'' \pm 0.0003''$ and can be obtained rapidly and inexpensively. The master used for grating replication was fabricated by HORIBA Jobin-Yvon (JY). This grating has a density of 5670 grooves/mm with parallel grooves and a sinusoidal profile. The grooves are created by etching the master substrate with photoresist exposed to a laser interference pattern.  This enables fabrication of high groove density onto a substrate of high optical quality.  To optimize packing geometry the graze angle is $4.4^{\circ}$ for the gratings. The gratings are replicated onto 104 x 104 mm substrates but these are subsequently laser cut to 20 mm in the groove dimension to ensure the desired resolution.  The specific cutting process utilizes femtosecond laser pulses that cut through the epoxy layer without raising its temperature (which would lead to layer delamination). After replication by JY, the gratings are coated with nickel for high reflectivity over the bandpass and to alleviate any bimetallic bending caused by the epoxy replication layer.

\begin{figure}
  \includegraphics[width=0.8\textwidth]{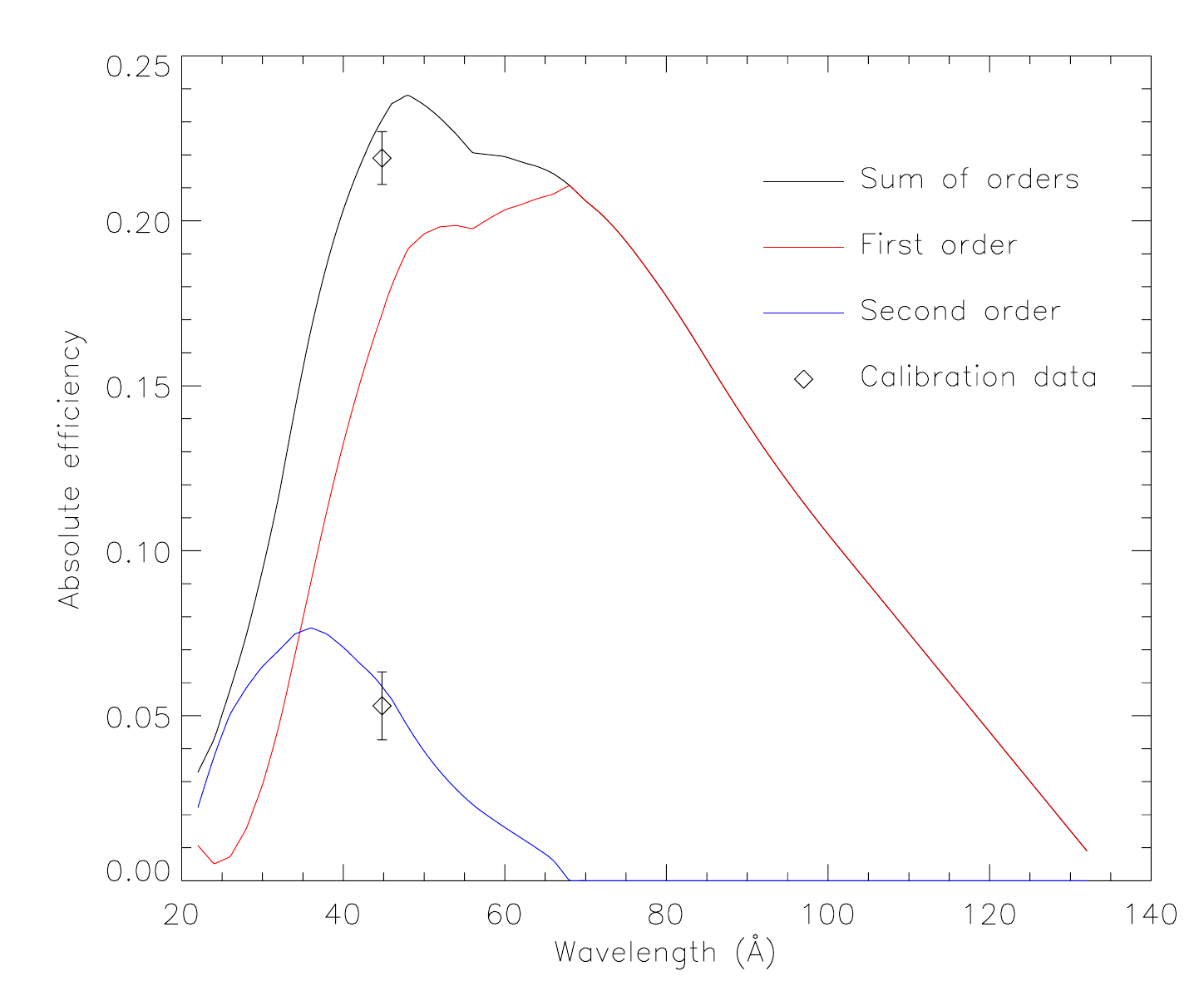}
\caption{Grating theoretical efficiency and calibration data. The calibration data shows the diffraction efficiency of a carbon emission line in both first and second order.}
\label{fig:7}      
\end{figure}

The gratings were tested for dispersion efficiency and matched theoretical predictions quite well (Figure \ref{fig:7}).  These theoretical predictions were calculated by the grating manufacturer, JY, using the actual groove profile obtained from atomic force microscopy.  Our empirical measurements are made using a Manson electron impact source fed monochromator producing a carbon K-$\alpha$ line at 0.28 keV.  The gratings are capable of placing 22\% of these photons into the positive first order spectrum with 5\% in positive second order.  

\begin{figure}
  \includegraphics[width=\textwidth]{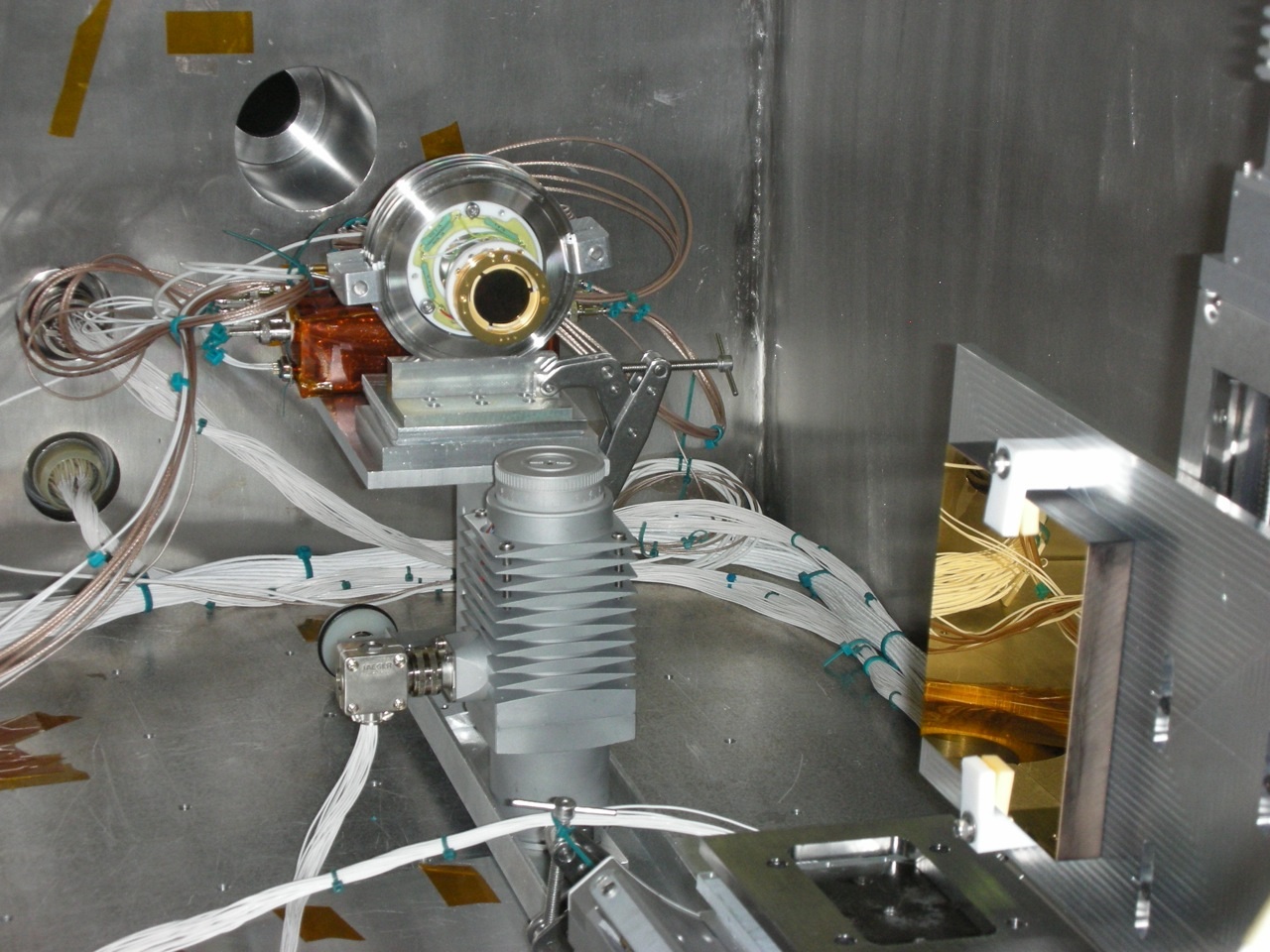}
\caption{Hardware setup for grating efficiency tests. Monochromatic X-rays are sent to the grating shown on the right, and are then dispersed via the geometry shown in Figure \ref{fig:5}. The detector (a micro-channel plate imager) is moved into the desired spectral line to observe the count rate. This count rate can be compared to the rate without the gratings in the beam to determine the efficiency of the gratings.}
\label{fig:8}      
\end{figure}

For future missions, the efficiency of these gratings and therefore the effective area of the spectrometer can be improved using several means. Modifying the profile of the grooves (known as blazing) can preferentially direct light into a preferred order. Lowering the graze angle from $4.4^{\circ}$ can shift efficiencies from lower energies to higher energies. Fortunately these do not greatly complicate our physical design. Figure \ref{fig:8} shows a hardware setup to test grating efficiencies at the University of Colorado.

\subsection{GEM Detectors}
	After approximately 2 meters of dispersion distance (throw) the spectral lines are recorded with Gaseous Electron Multiplier (GEM) detectors. These detectors were chosen to provide an inexpensive means of obtaining a large format (10 x 10 cm) necessary to capture our desired bandpass. The GEM detector has a thin (5000 \AA) window made of polyimide and carbon through which X-ray photons pass to enter the detector. Once inside the GEM, the photons ionize the argon gas in the drift region between the window and the first GEM foil. The ionization energy of argon is 26 eV and thus soft X-rays in our bandpass create $\sim5-30$ ion-electron pairs. The drift region is approximately 5mm in depth, enough for high probability of interaction and minimal electron cloud spreading which would reduce our resolution.

\begin{figure}
  \includegraphics[width=\textwidth]{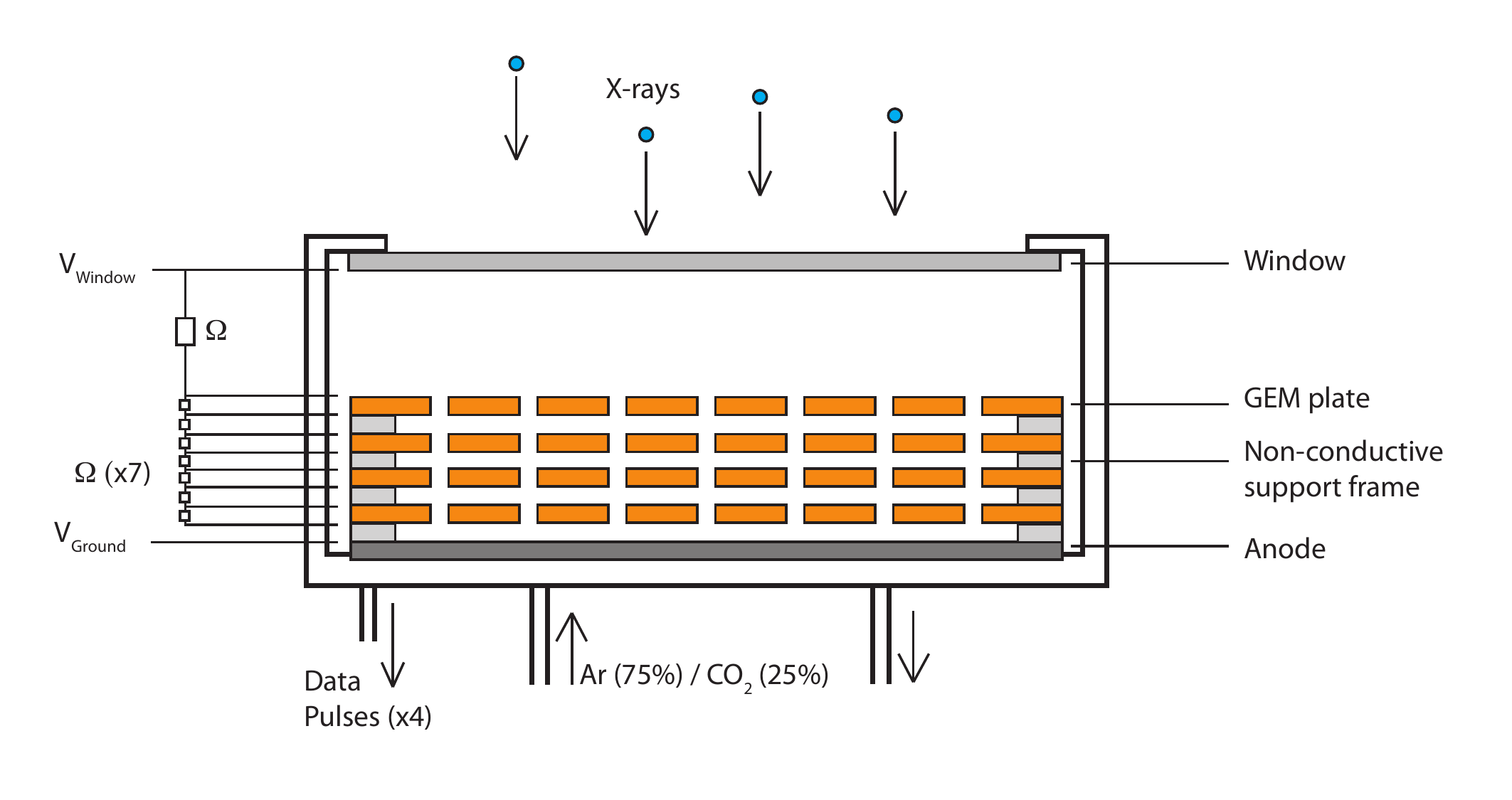}
\caption{Schematic of a GEM detector. Light enters from above through a thin polyimide / carbon window. The X-rays have approximately 5 mm in the drift region before the first GEM foil to ionize the argon gas. The window is held at a high negative voltage, typically 4000 Volts. The voltage drop to the top of the first gem plate is 500-700 volts. Each foil has a potential drop of $\sim400$ Volts from top to bottom (there is a 100 micron thick insulator between conductive copper layers). A potential drop of $\sim200$ Volts is established in the 1mm gap between plates. The anode at the bottom is held at ground voltage.}
\label{fig:9}      
\end{figure}

\begin{figure}
  \includegraphics[width=0.5\textwidth]{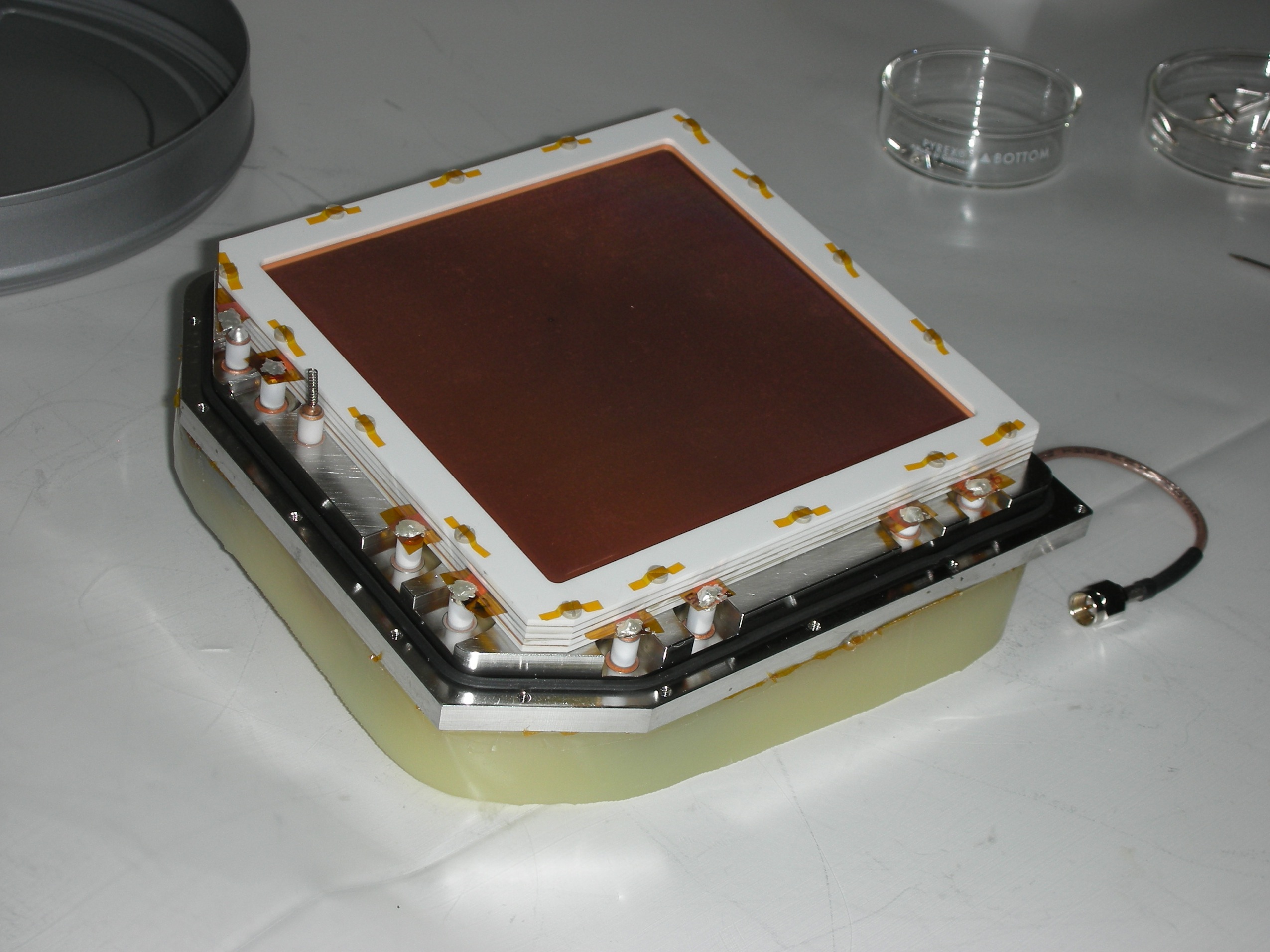}
  \includegraphics[width=0.5\textwidth]{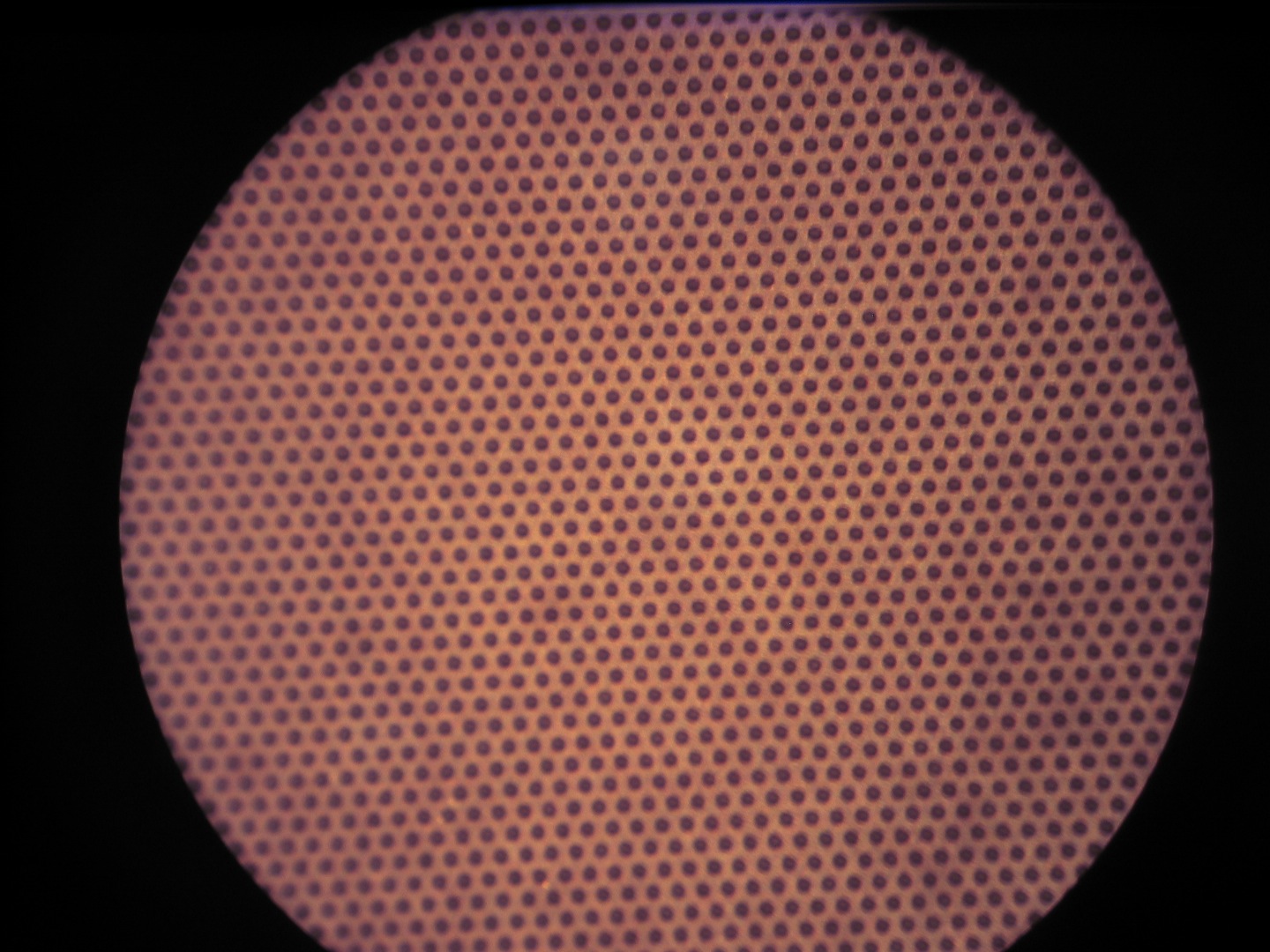}
\caption{Left - The internal assembly of our GEM detectors.  Includes 4 perforated GEM foils with base of 100 micron thickness LCP and a coating of 8 micron thick copper on each side for conductivity.  The foils are laser etched to form holes with a 140 micron pitch and 70 micron diameter. Right - A 7x magnified view of a CyXESS GEM plate showing the pores.}
\label{fig:10}      
\end{figure}

The GEM detectors have a series of four porous foils encased in an Ar/CO$_2$ gas chamber held at 14.5 psi.  The argon acts as the source of electrons as X-rays ionize the gas. The CO$_2$ acts as a quenching agent and neutralizes the ionized argon via charge exchange. A schematic of GEM operation is shown in Figure \ref{fig:9} and a partially disassembled GEM is shown in Figure \ref{fig:10}. The window itself is made of polyimide, but the underside is coated with a 300\AA\  layer of carbon for conductivity. This window is held at a high negative voltage, while the top of the first GEM plate is held at a slightly lower negative voltage. The electric field thus directs photons downward through the drift region. The GEM foil itself is nonconductive liquid crystal polymer (LCP), 100 microns thick with an 8 micron thick coating of conductive copper on both surfaces. These two surfaces are also held at different voltages, so that as the electrons pass through the pores, they experience a concentrated electric field and voltage gradient from one side of the plate to the other, causing further collisions, an electron cascade and amplified signal. The voltage drop within a pore is approximately 400 Volts. This cascade is repeated at each of the four GEM foils, providing the necessary gain to detect soft X-rays which only liberate a few initial ion-electron pairs.

The GEM foils are thin and must be mounted in a fashion to prevent large scale motions, particularly during launch. This is achieved by heating the GEM foils to $50^{\circ}$ C and allowing them to expand. Ceramic frames are then epoxied to the GEM foil while hot. After curing, the assembly is allowed to cool, thus the contracting GEM foils become taut in their frames.

The anode, located at the bottom of the detector, is held at ground and collects the electron cloud. The 100 mm x 100 mm anode is a serpentine cross delay line made of palladium silver on an alumina substrate. The distance between parallel lines on the weave is 0.57 mm. The x-axis serpentine line is separated from the y-axis serpentine line by a dielectric compound. The charge cloud is measured at the end of each axis by the detector electronics. The time delay in arrival from each end of an axis is translated into a physical position on the anode. The size of the overall charge is also proportional to the energy of the original photon, giving some energy sensitivity. These detectors were challenging to work with for the CyXESS mission.  A high gain ($10^4 - 10^5$) is necessary in order to amplify soft (1/4 keV) X-rays, but is difficult to maintain.  The GEM foils had difficulty sustaining the voltage drop from one side of the plate to the other due to manufacturing defects in the pores. Irregularity in the distance between the Cu plating on each side due to a badly shaped pore, or a minute piece of Cu extending into the pore can lead to a short across the plate.  The shorts in these bad pores result in hot spots on the detector image.  They also short out the voltage drop across the plate, decreasing the overall gain of the detector and its sensitivity to soft X-rays.  We have investigated other manufacturing techniques to remedy these problems and now use plates fabricated by a new manufacturer, SciEnergy (see \cite{Tamagawa2006}, \cite{Tamagawa2008} \cite{Tamagawa2009} and references therein). These plates are laser cut rather than mechanically cut or chemically etched, producing high fidelity pores and allowing the plates to sustain the required voltages. They display substantially less anomalous behavior, allowing observations with substantially less noise and fewer breakdown events, and higher sensitivity. Additionally these plates require little to no warmup time, whereas the chemically etched plates would only perform optimally after more than an hour of use.

\begin{figure}
  \includegraphics[height=1.8in]{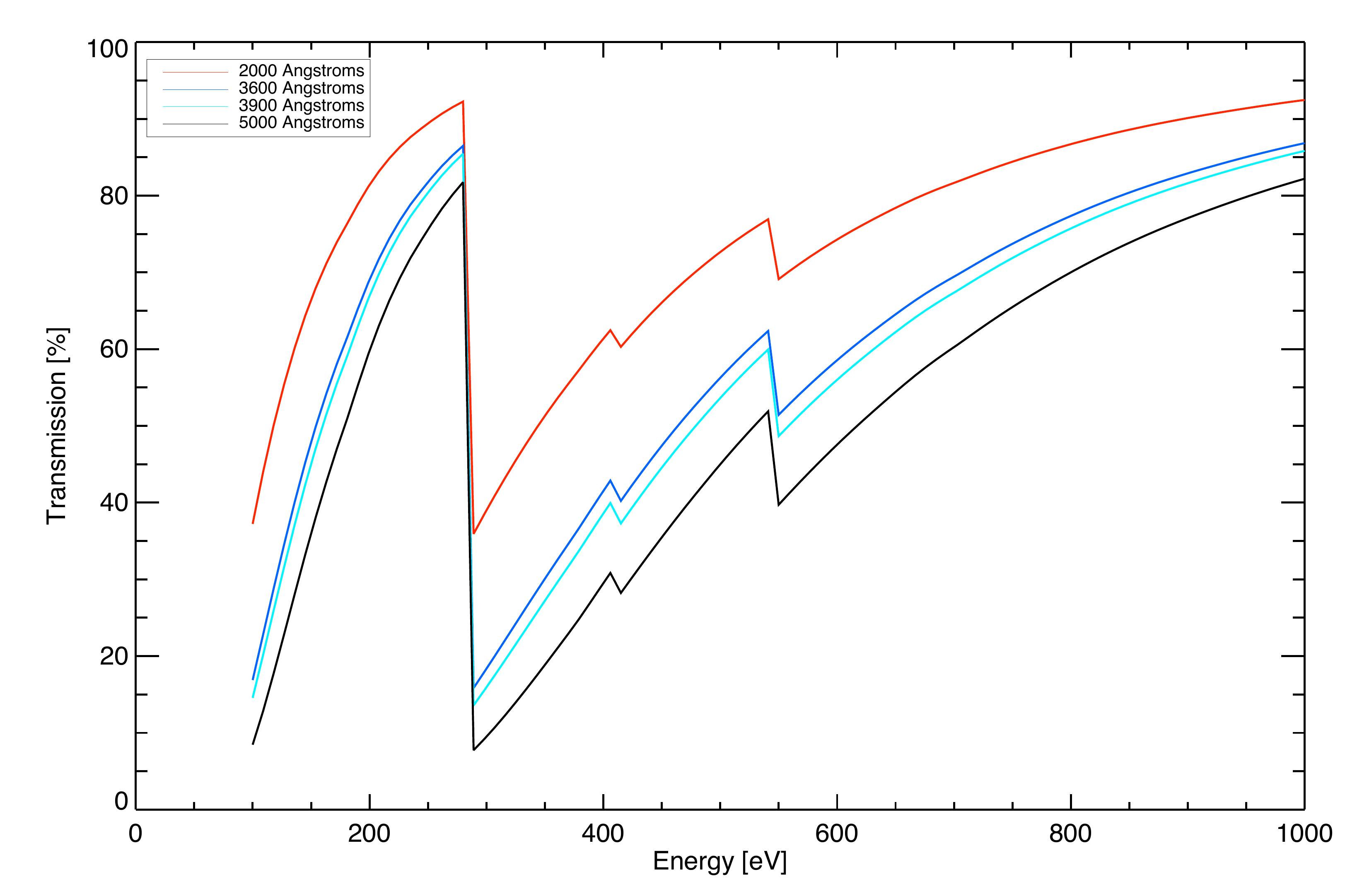}
  \includegraphics[height=1.8in]{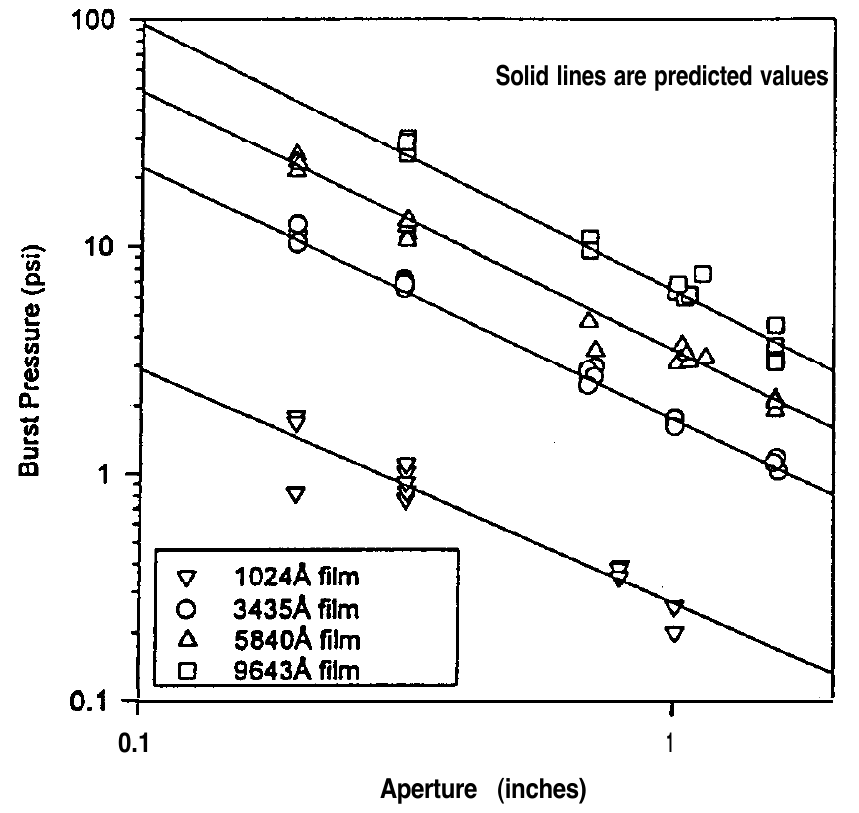}
\caption{Left - Transmission as a function of thickness for the polyimide layer of the GEM windows. Right - Strength versus aperture size of polyimide. These windows need to sustain at least 14.5 psi for normal operation. Typically a safety factor of $\ge3$ is desired between operating pressure and burst pressure. The aperture for EXOS is 0.05 and the thickness is 5000 Angstroms.}
\label{fig:11}      
\end{figure}

We also replaced the windows on these detectors with a slightly thicker design. The initial windows were 3600-3900 \AA\ thick, which withstood the pressure differential (14.5 psi inside against an evacuated payload) adequately during testing. However, a hole occurred in one of the windows during the CyXESS flight, causing a partial loss of functionality. To prevent this from reoccurring we obtained new windows from Luxel with 5000 \AA\ thick polyimide. Interior to this is a 300 \AA\ thick layer of carbon for conductivity. This film is supported by a 20 lines/inch stainless steel mesh. We expect $\sim10\%$ loss in transmission due to the increased thickness. Extrapolation from data taken by Luxel with larger apertures gives nearly 100\% increase in strength from this $\sim30\%$ increase in thickness (Figure \ref{fig:11}).

Due to the window size and thinness, it is difficult to prevent minor leakage of the detector gas into the payload. As the detector is sensitive to changes in pressure on the level of $\sim1\%$, this leakage is a serious concern. The detectors therefore have an on-board gas system housed within the electronics section behind the detector bulkhead. This gas system consists of a gas reservoir, a regulator for rough pressure stabilization, and a proportional valve to establish the detector pressure to an accuracy of $\sim0.1\%$. This proportional valve also allows for real time monitoring of detector pressure during testing and flight. This diagnostic provides another means of assessing detector performance.

\begin{figure}
  \includegraphics[width=0.8\textwidth]{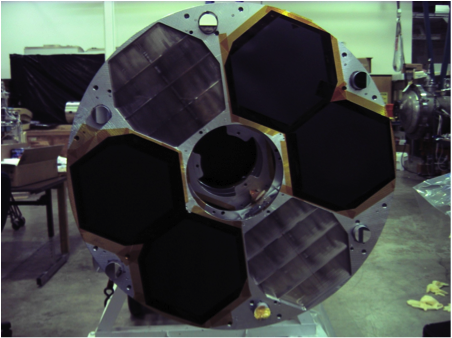}
\caption{Telescope apertures. Two modules are fully filled, while up to 4 additional modules can be added for future flights. Currently the unused modules have been baffled with black kapton MTB series from DuPont.}
\label{fig:12}      
\end{figure}

\begin{figure}
  \includegraphics[width=\textwidth]{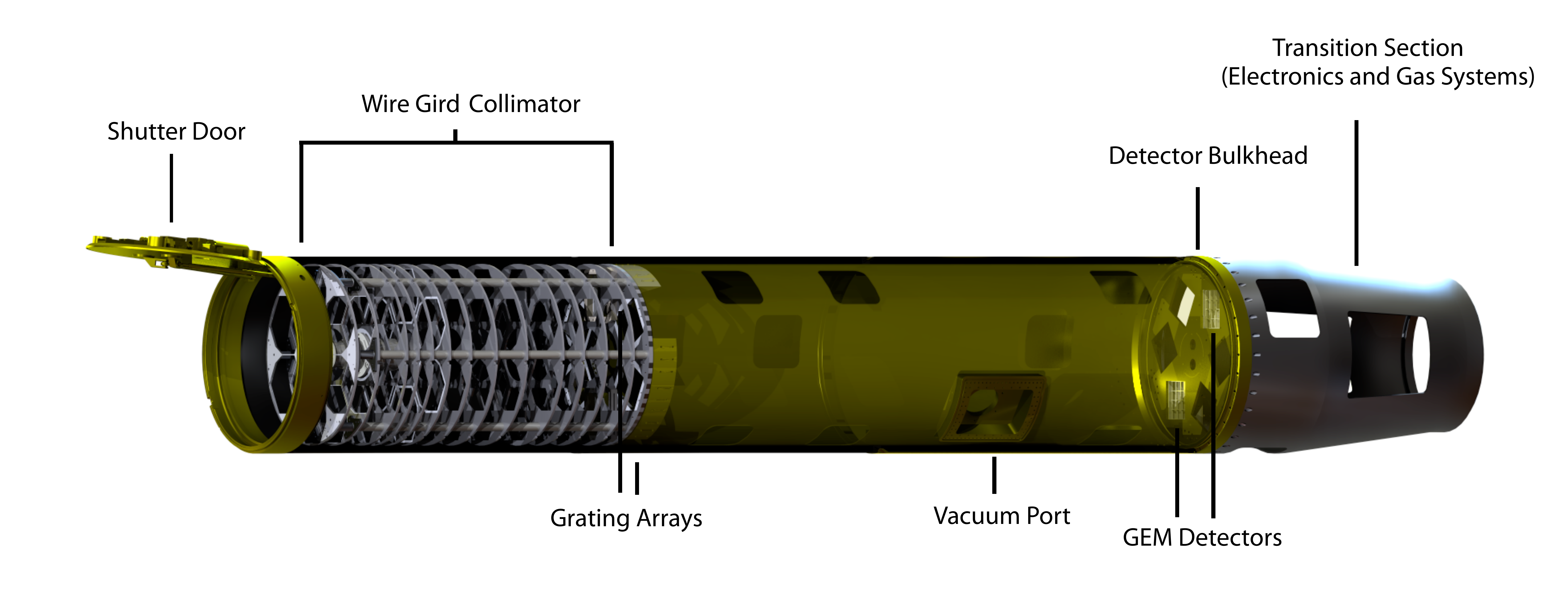}
  \includegraphics[width=\textwidth]{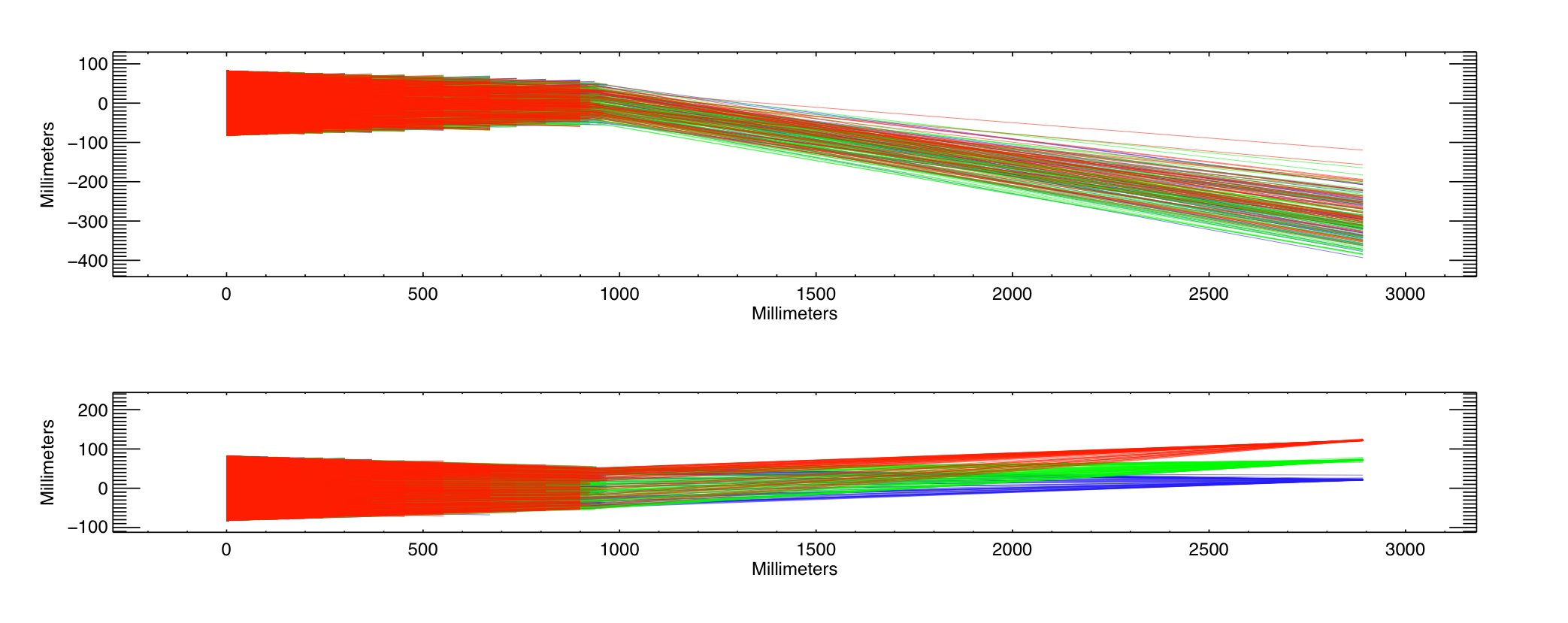}
\caption{Schematic / raytrace of the instrument. Top image shows a photo rendering of the entire payload. The middle plot shows the raytrace forming a line along the direction of the wire grid slits, while the bottom image shows the photons being dispersed. The first meter is occupied by the wire-grid collimator with the gratings placed immediately after. The spectral lines are dispersed over the remaining 2 meters and show up on the detector as vertical lines.}
\label{fig:13}      
\end{figure}

The payload was initially designed to include 6 modules within a 22-inch rocket skin.  Given a limited budget on the CyXESS flight, we filled only 2 of these modules.  For EXOS we flew these 2 modules again (see Figure \ref{fig:12}) to show the full capabilities of our refurbished detectors. For the next flight, we will again fly 2 modules with our third generation of GEM detectors and additional improvements to the wire-grid collimator. The full payload is shown in Figure \ref{fig:13} as a model rendering along with a raytrace of the entire optical system. 

\section{Pre-Flight Calibration Data}
\label{sec:2}
\begin{figure}
  \includegraphics[width=0.8\textwidth]{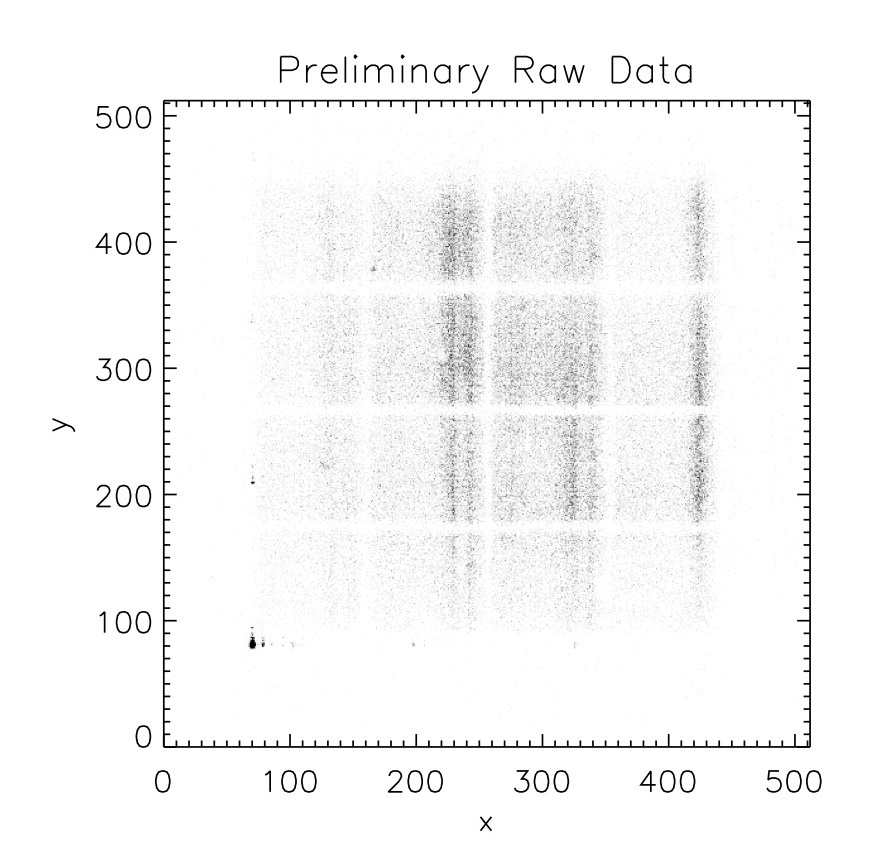}
\caption{Lab calibration data.  The vertical lines of counts represent spectral lines from a source emitting primarily carbon and oxygen emission lines. The bright spot in the lower left corner is the detector stim pulse. These lines match the raytrace, showing a FWHM of $\sim2$ mm.}
\label{fig:14}      
\end{figure}

The EXOS sounding rocket underwent final assembly and calibration in the summer of 2009. An example of calibration data is shown in Figure \ref{fig:14}. These data were obtained by placing the payload in the Rocket Calibration Facility (RCF) at the University of Colorado.  The RCF is a $30^{\prime}$ long vacuum chamber that is $30^{\prime\prime}$ in diameter and designed for full-system calibrations of suborbital rocket payloads (which are typically $<3$ meters in length and $17^{\prime\prime}$ or $22^{\prime\prime}$ in diameter). This facility is capable of vacuum levels $\sim 10^{-7}$ torr and is capable of utilizing a variety of light sources and motion apparatus. For this payload the light source was an electron impact X-ray source. This source emits approximately like a point source. The emission comes from a spot $\sim300 \mu$m in diameter and is windowed to a cone of emission $\sim20^{\circ}$ width. Though not an extended light source, it can be moved vertically and horizontally during an exposure to simulate a larger object and fill the EXOS field of view. The 4x4 grid pattern on the detector face is caused by the cast shadow of the aluminum window frame. This frame supports the thin polyimide mesh and has 3 cross bars along both dimensions for support. The small dots in the lower left corner of the image is a signal, known as the stim pulse, generated by the detector electronics at $\sim10$ Hz that is used to verify detector functionality. This stim pulse verifies that the detector electronics (timing to digital converter, amplifiers, etc.) are properly connected to the detectors, and that the data is being properly passed through the payload, telemetry and ground support electronics. This pulse is also used to insure that the data extraction and analysis software is properly handling the detector data. The low level signal over the entire detector face is the X-ray continuum emission from our source. The detectors have a well-characterized stable background count rate of 2 cts/s. 

\begin{figure}
  \includegraphics[width=0.5\textwidth]{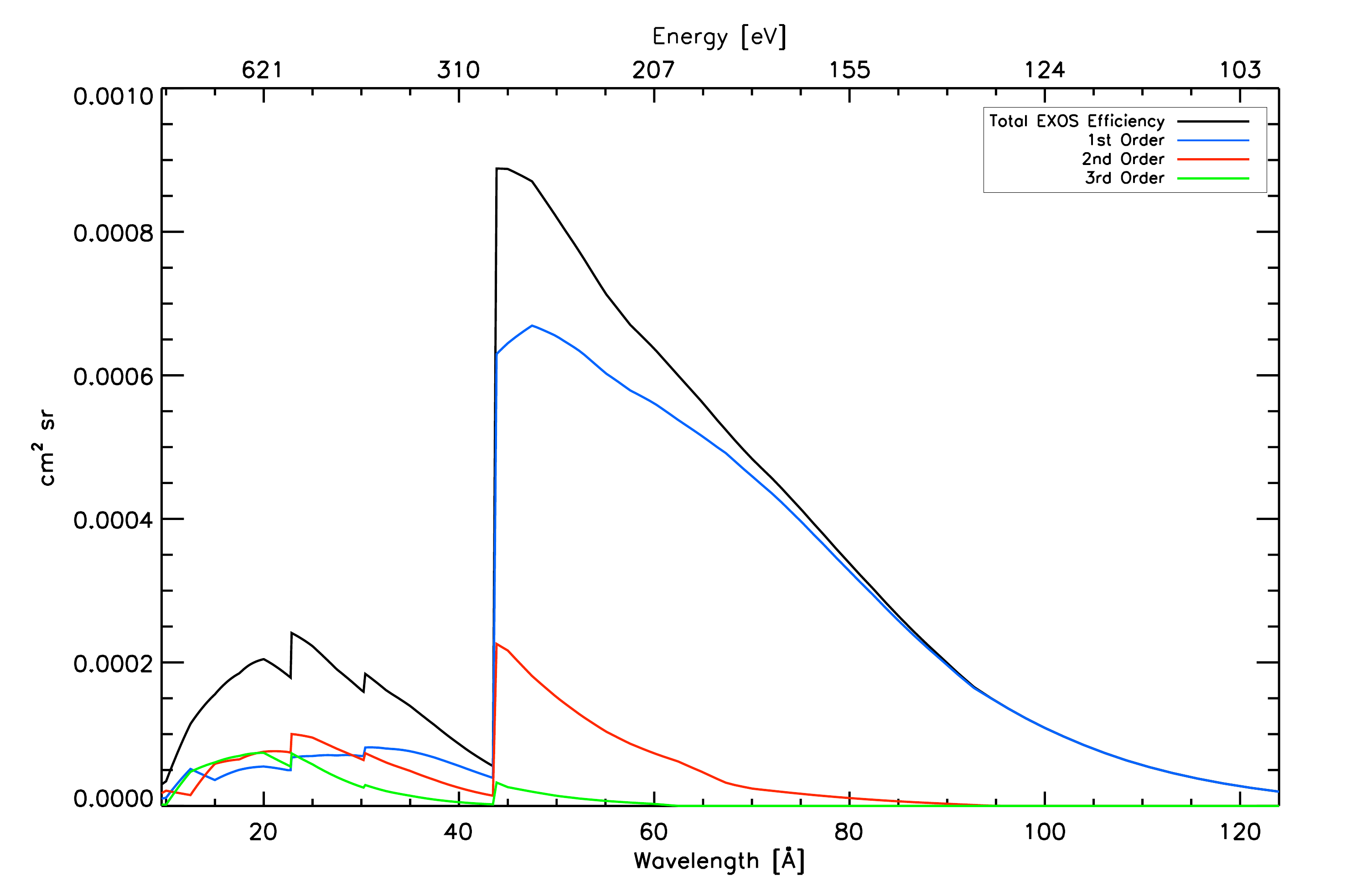}
  \includegraphics[width=0.5\textwidth]{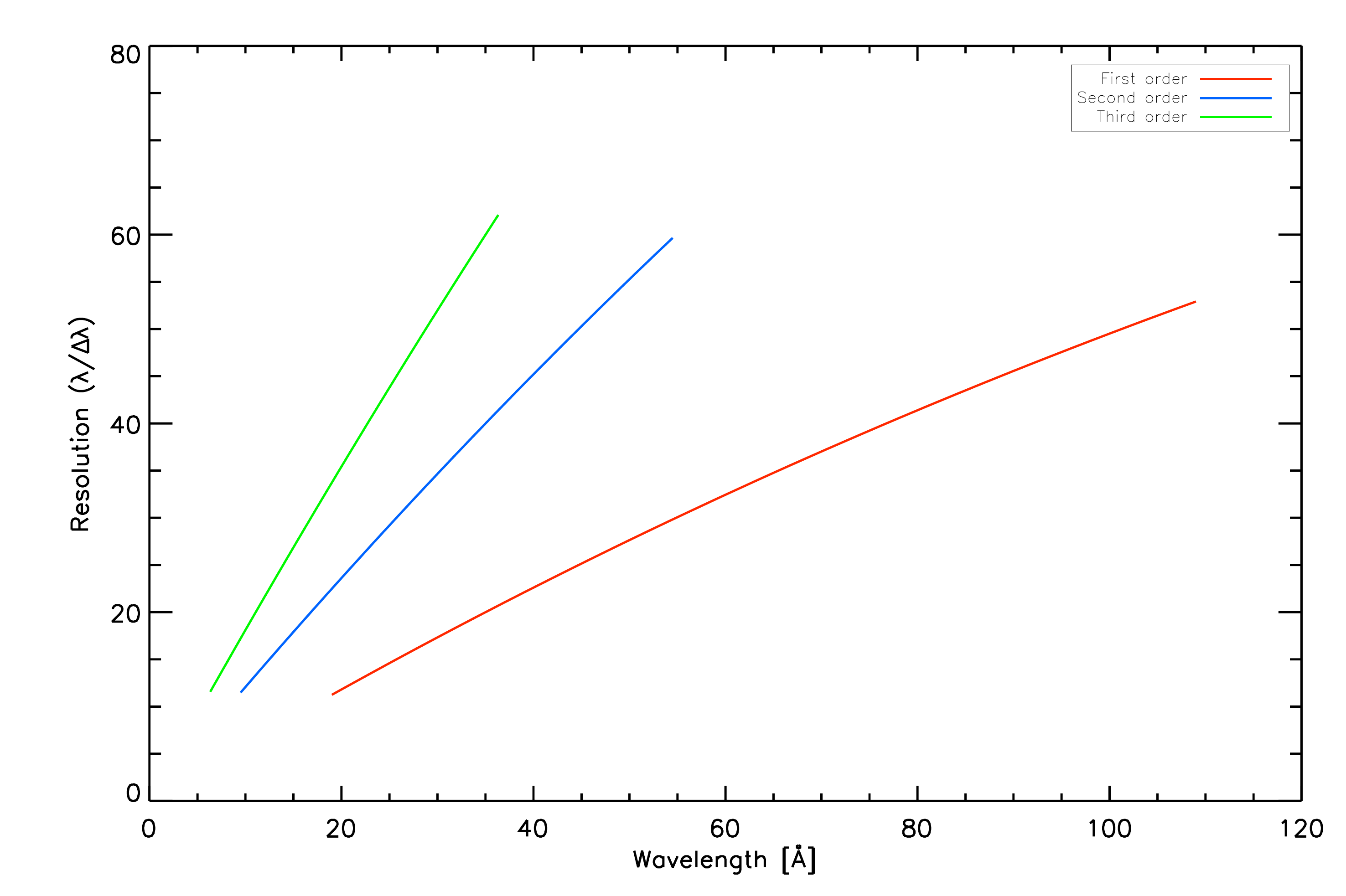}
\caption{Left - Theoretical grasp curve for the telescope. This shape assumes the target flux is distributed in a ring shape of $3.25^\circ$ outer diameter and $3^\circ$ inner diameter. Experience with these types of gratings indicates that the peak efficiency (due to the pseudo sinusoidal blaze) is likely much wider than theory predicts. The sharp cut off at 44 \AA\ is due to the carbon edge in our polyimide carbon window. Right - Theoretical resolution of the system as a function of wavelength. The resolution deviates from linear towards longer wavelengths (farther from zero order) due to the different path lengths travelled from either side of the grating array}
\label{fig:15}      
\end{figure}

The observed spectral lines match the raytrace well. They show FWHM of $\sim2$mm as anticipated and have throughput similar to or higher than CyXESS calibration data. The detector behavior is also greatly improved. They show virtually no hot spots or other undesired behavior. The throughput also increased due to both our improved gain on the new GEM foils as well as the replacement of several of our wire-grid plates that were damaged during integration prior to the flight of CyXESS. The grasp of the payload is an important, but exceedingly difficult metric to determine. Grasp is measured in cm$^2\cdot$steradians$\cdot$seconds. Since observing time is extremely limited in a sounding rocket launch, the instrument must maximize both area and FOV. The FOV is defined by the geometry of the collimator structure, while the area is the convolution of the slit size and the wavelength dependent efficiencies of the gratings and detector. The primary difficultly is that every recorded count can originate from a range of graze angles ($\gamma$) on the gratings and not every photon diffracted by the gratings strikes the detector face (the spectral lines are longer than the detector is tall). Graze angles from $\sim2^{\circ}-6^{\circ}$ are seen by the detector (though not evenly distributed). By combining the raytrace with diffraction efficiency curves calculated by JY we can calculate an approximation to the instrument's grasp, as shown in Figure \ref{fig:15}. This curve is modified by the spatial distribution of the source target (an annular ring was used in this simulation). We are in the process of developing an extended X-ray source for calibration purposes that will allow a more empirical assessment of our effective area. The development of this new source will also allow better wavelength calibration and could be implemented as an onboard calibration source for future flights.

\section{Launch Results}
\begin{figure}
  \includegraphics[width=\textwidth]{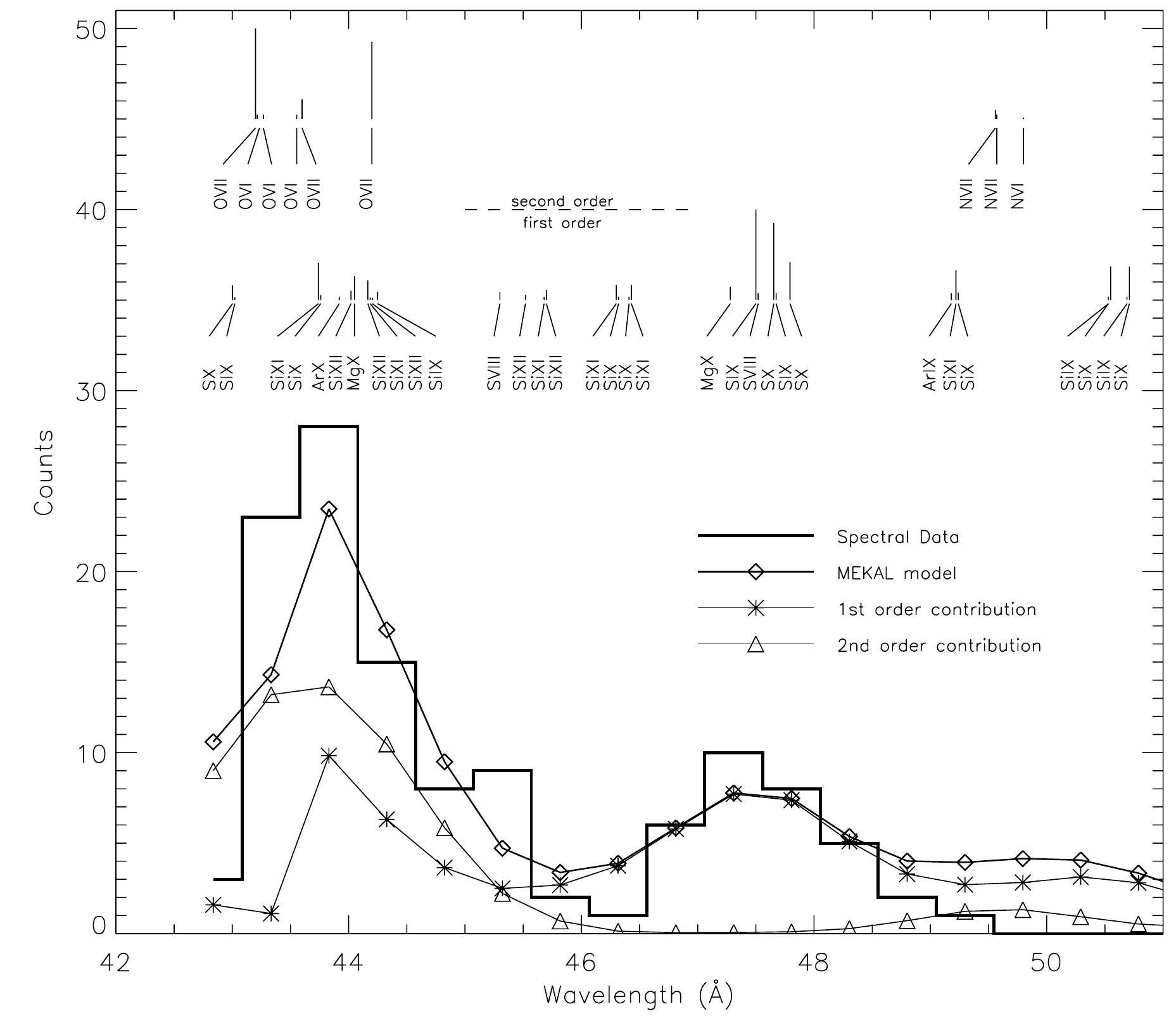}
\caption{Spectrum taken by the CyXESS payload.}
\label{fig:16}      
\end{figure}

The CyXESS suborbital rocket was launched on November 20, 2006 (flight 36.224) from White Sands Missile Range (WSMR) at 7:00pm (MST). Data were recorded for 345 seconds of the flight. Unfortunately, a large breakdown event occurred at the beginning of the flight when the detector's high voltage interacted with residual gas inside the payload. The payload is evacuated through the vacuum port (Figure \ref{fig:13}) prior to launch, but unfortunately the vacuum pump must be removed $\sim1-2$ hours prior to launch. This timeline is driven by the safety concerns of arming the rocket motors. This delay between pumping and launch allows for a significant amount of gas to build up in the payload primarily from minor leaks in our detector windows and outgassing of the rocket skins. This breakdown event rendered one detector useless and left the other detector noisy. Useable data were obtained only from the final 65 seconds of flight. The resultant spectrum is shown in Figure \ref{fig:16}. This spectrum shows two features dominated by O VII, Si XI, Si XII, and Mg X around 44 \AA, and S IX and S X around 47 \AA. Fits to this spectrum give an equilibrium plasma at kT$_e$ = 0.14 keV and an observed depletion of Si, likely indicating the presence of dust in the form of silicate grains (\cite{McEntaffer2008}).

Due to the brevity of useable data from the CyXESS flight, the EXOS suborbital rocket was launched  (flight 36.252) on the same target. This launch occurred on November 13, 2009 at 7:30 PM (MST) again at WSMR. The target was acquired 106 seconds after launch (almost simultaneous with detector HV turn on) and detector diagnostics indicate that there were no telemetry or electronics problems. Data was recorded for 364 seconds during flight. During initial HV turn on, the pressure in the payload was slightly too high ($>10^{-4}$ Torr) and caused a discharge event in front of the detector window. Fortunately as the residual material pumped into space, this event quickly went away and only cost 4\% of our observation time. Our discharge event was much shorter than that of the CyXESS flight due to a shorter pump to launch delay, thicker detector windows, and an improved gas system. This type of event is easy to diagnose and remove due to the higher count rates and vastly different pulse height distribution in comparsion to a soft X-ray source. The other possible type of detector noise, localized hotspots, was also seen. These hotspots were seen sporadically in flight, but were small in physical size (approximately 1 cm in diameter in comparison to the 100 cm$^2$ GEM) and caused nothing more than a slight drop (on the order of several percent) in effective area of a few spectral bins. This noise is simple to diagnose and remove from the data given its small spatial footprint. Considering the totality of photons lost from the discharge event and hotpot activity, we calculate that we collected useable data for over 90\% of flight time. The onboard diagnostics of gas pressure, HV level, power, etc. were all nominal during flight, indicating optimal instrument performance.  

\begin{figure}
  \includegraphics[width=\textwidth]{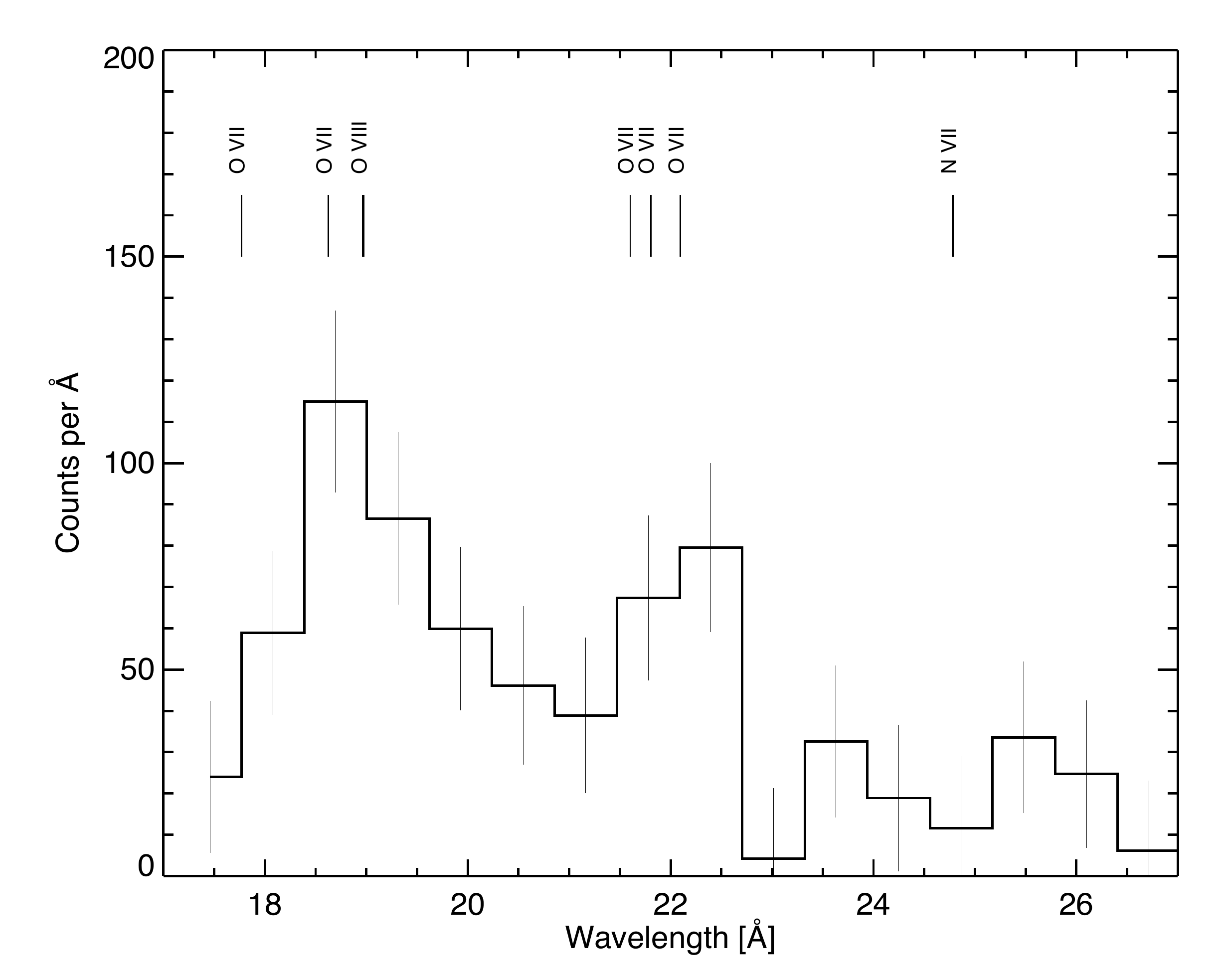}
\caption{Preliminary results of the EXOS payload. Line identifications are based on predicted transitions of thermal plasma models.}
\label{fig:17}      
\end{figure}

Due to the hard landing the payload was damaged in several places, including the rocket skins, the collimator structure and the gas system. Unfortunately this prevents us from conducting post-flight calibrations on the payload. A partial results spectrum is shown in Figure \ref{fig:17}. This spectrum shows likely oxygen emission lines at $\sim19$\AA\ and 22\AA\ and nitrogen at $\sim25$ \AA. Expected transitions based on thermal plasma models (\cite{Borkowski2001}, \cite{Hamilton1983}, \cite{Borkowski1994}, \cite{Liedahl1995}) are labelled. Analysis of the full spectrum for publication is currently underway. 

\section{Launch Schedule}
Our intention is to launch approximately once a year for 4 years. Our first launch was to prove the full capability of our instrument on the Cygnus Loop. Though the previous flight of CyXESS was a success, the difficulties with our detectors (due to both the noisy GEM plates and a torn polyimide window) prevented us from truly showcasing the full abilities of the instrument. After the recent success of EXOS,  we now intend to observe the Vela SNR for our third flight that is scheduled for a Spring 2011 launch from WSMR. After this flight, we intend to add 2 additional GEM detectors to capture the photons diffracted into the negative spectral orders. This will double our effective area without having to obtain and align any additional optics. These improvements will allow us to achieve our primary science goal for this payload - observations of the diffuse soft X-ray background. Existing observations with the DXS (\cite{Sanders1998}, \cite{Sanders2001}) and XQC (\cite{McCammon2002}) instruments have resulted in a multitude of unresolved line blends. The authors are unable to adequately fit physical models to the data and attribute this to a lack of resolution ($\sim20-40 \; E/\Delta E$ in the 1/4 keV bandpass). EXOS currently achieves resolutions of $\sim60 \; (\lambda / \Delta \lambda)$ and will increase to $\sim100$ when reconfigured for this flight. This will allow more detailed observations and lead to better line identification and physical model analysis.

While some of this emission may be due to a local hot bubble (LHB) of interstellar gas in our local galactic neighborhood, much of the flux can be attributed to an interaction between the solar wind and local neutrals (\cite{Lisse1996}, \cite{Cravens1997}, \cite{Cravens2001}).  Ions in the solar wind can undergo charge exchange with these neutrals in the heliosphere and Earth's geocorona. We hope to begin to separate this charge exchange emission that currently contaminates the LHB emission by observing twice under differing solar wind conditions. Because the solar wind charge exchange (SWCX) emission occurs even when the sun is quiescent, it is not possible to derive the true LHB emission without fully understanding the variability of SWCX. By observing which emission lines vary with the solar wind, we will be able to determine which spectral lines are caused by charge exchange versus plasma lines at the LHB boundary. This technique has been successfully proven by \cite{Snowden2004} and \cite{Fujimoto2007} at higher energies. As most of the SWXC occurs at lower energies (around 1/4 keV), we anticipate a wealth of identified lines with our instrument. From here we can begin to better understand the LHB with a cleaner soft X-ray sectrum.

This payload is also an ideal candidate for a platform that permits longer observations, such as orbital sounding rockets. Due to our large FOV, our target acquisition and pointing requirements are only on the order of arc minutes. This is a relatively simple task considering rocket star trackers are currently capable of arc second level pointing. Furthermore, we have already identified a plethora of individual targets from which we could extract unique science. In addition, our previously mentioned charge exchange experiment can be accomplished much more robustly with longer observations. A longer observation will be able to take into account subtleties such as solar wind compositional changes, differences between heliospheric and geocorona charge-exchange, and better mapping of latitudinal influences. 

Because each module is independent, we have the flexibility to design each module to maximize scientific return. If particular science goals require a certain resolution, bandpass or FOV then a subset of modules can be specifically designed for this purpose. For example: a finer collimator slit width would provide better spectral resolution, a different grating design would shift our effective area to a different bandpass and a larger opening angle would provide us with a larger FOV. These alterations provide both a better scientific return as well as proof of concept for optimizing our optics and detector technologies for future missions. Different focusing optics or detectors (such as CCDs) could also be swapped into the design to provide flight experience for different technologies. Modular independence also provides a highly valuable risk mitigation advantage. Since each module operates independently, a failure in one or more detectors will still yield scientifically compelling results and mission success. 

\section{Summary}
The University of Colorado, Boulder has designed a payload capable of high resolution diffuse spectroscopy in the soft X-ray (17-107 \AA) bandpass. The payload's optical path is defined by its three major components: a wire-grid collimator, an off-plane grating array and GEM detectors. This payload has been launched twice (as CyXESS and EXOS) on the Cygnus Loop and is scheduled (as CODEX) for an early 2011 launch on the Vela SNR. We plan to install more modules over the next 3 years and launch the payload approximately once a year for a total of 3 additional launches, including the upcoming CODEX launch. The wide range of potential astronomical observations allows us to obtain a strong science return on supernova remnants, galactic halo emission, the local hot bubble and solar wind charge exchange with any number of modules filled. Given the opportunity, this payload has strong potential for longer duration observations due to its unique observational capabilities, loose pointing requirements, abundance of scientific targets and flexibility due to its modular design.

\section{Acknowledgements}
This project was funded by NASA grant NNX09AC23G. We thank the Wallops and White Sands Missile Range support personnel from flights 36.224 and 36.252 for all of their critical work on the project. We would also like to thank Luxel for their help in improving our GEM windows, Toru Tamagawa for his assistance with our GEM plates, Charlie Zabel for his constant help with our gas system and the Hodgsons for cutting our metal. The success of these missions was reliant on the work done by Travis Curtis, Mike Kaiser, Nico Nell, Eric Schindhelm, Ted Schultz, Ann Shipley and Ben Zeiger.

\clearpage

\begin{table}[htdp]
\caption{Parameters for the EXOS instrument}
\begin{center}
\begin{tabular}{l@{\extracolsep{.7cm}}c@{\extracolsep{1.5cm}}l}
Parameter & System Component & Value \\
\hline
\hline
&Payload \\
\hline
Optical Path Length && 3 m \\
Payload Diameter && 22 inches \\
Observing Time && 364 seconds \\
Field of View && $3.25^{\circ} \times 3.25^{\circ}$ \\
Necessary Pointing Accuracy && $5^{\prime}$ \\
Bandpass && 17 - 107 \AA\ \\
Line size && 1.7 mm (wide) $\times$ \\
&& 100 mm (tall) \\
Resolution && 10 - 60 $(\lambda / \Delta \lambda)$ \\
\hline
&Collimator \\
\hline
Collimator Length && 1 meter \\
Number of Wire-grid Plates && 24 per module \\
Entrance Slit Width && 725 microns \\
Final Slit Width && 500 microns \\
\hline
&Gratings \\
\hline
Grating Size && 100 mm $\times$ 20 mm \\
Number of Gratings && 67 per module \\
Groove Profile && sinusoidal \\
Groove Density && 5670 groves / mm \\ 
Grating Coating && Nickel \\
Dispersion Distance (Throw) && 2 meters \\
\hline
&GEM detectors \\
\hline
Detector Size && 100 mm $\times$ 100 mm \\
Detector Voltage && $\sim4000$ Volts \\
Detector Gas && Ar/CO$_2 \; (75\% / 25 \%)$ \\
Detector Pressure && 14.5 psi \\
Detector Spatial Resolution && $\sim100-200 \mu$m \\
\hline
\end{tabular}
\end{center}
\label{table:exos_parameters}
\end{table}%

\clearpage

\clearpage

\clearpage

\clearpage

\clearpage

\clearpage

\clearpage

\clearpage

\clearpage

\clearpage

\clearpage

\clearpage

\clearpage

\clearpage

\clearpage

\clearpage

\clearpage


\begin{thebibliography}{}
\bibitem[Blair et al. (1999)]{Blair1999} Blair, W P, Ravi Sankrit, John C Raymond, and Knox S Long. Distance to the Cygnus Loop from the Hubble Space Telescope Imaging of the Primary Shock Front. The Astronomical Journal 118 (1999): 942-947.
\bibitem[Borkowski et al. (1994)]{Borkowski1994} Borkowski, K J, C L Sarazin, and J M Blondin. On the X-ray Spectrum of Kepler's Supernova Remnant. The Astrophysical Journal 429 (1994): 710-725.
\bibitem[Borkowski et al. (2001)]{Borkowski2001} Borkowski, Kazimierz J., William J Lyerly, and S P Reynolds. Supernova Remnants in the Sedov Expanion Phase: Thermal X-ray Emission. The Astrophysical Journal 548 (2001): 820-835.
\bibitem[Casement et al. (2010)]{Casement2010} Casement, S, R L McEntaffer, W Cash, T Johnson, C Lillie, and D Dailey. A tower concept for the off-plane x-ray grating spectrometer for the International X-ray Observatory. Instrumentation 7732 (2010): 77323W-77323W-8.
\bibitem[Cash (1982)]{Cash1982} Cash, W. Echelle spectrographs at grazing incidence. Applied Optics 21, no. 4 (1982): 710-717.
\bibitem[Cash (1991)]{Cash1991} Cash, W. X-ray optics: a technique for high resolution imaging. Applied Pptics 30 (July 1991) 1749-1759.
\bibitem[Chandra Proposers' Observatory Guide (2009)]{Chandra2009} Chandra X-ray Center. The Chandra Proposers' Observatory Guide. Chandra Project 12 (2009).
\bibitem[Cox (2005)]{Cox2005} Cox, Donald P. The Three-Phase Interstellar Medium Revisited. Annual Review of Astronomy and Astrophysics 43, no. 1 (September 2005): 337-385.
\bibitem[Cravens (1997)]{Cravens1997} Cravens, TE. Comet Hyakutake X-ray Source: Charge Transfer of Solar Wind Heavy Ions. Geophysical Research Letters 24, no. 1 (1997): 105108.
\bibitem[Cravens et al. (2001)]{Cravens2001} Cravens, T.E., I.P. Robertson, and S L Snowden. Temporal variation of geocoronal and heliospheric X-ray emission associated with the solar wind interaction with neutrals. JGR 106, no. A11 (2001): 24883-24892.
\bibitem[Flanagan et al. (2004)]{Flanagan2004} Flanagan, K A, C. R. Canizares, D. Dewey, J. C. Houck, A. C. Fredericks, M. L. Schattenburg, T. H. Markert, and D. S. Davis. Chandra High Resolution X-Ray Spectrum of Supernova Remnant 1E 0102.27219. The Astrophysical Journal 605, no. 1 (April 2004): 230-246.
\bibitem[Fujimoto et al. (2007)]{Fujimoto2007} Fujimoto, R, K Mitsuda, D McCammon, Y Takei, M Bauer, Y Ishisaki, F S Porter, H Yamaguchi, K Hayashida, and N Y Yamasaki. Evidence for Solar-Wind Charge-Exchange X-Ray Emission from the Earth's Magnetosheath. Publ. Astron. Soc. Japan 59 (2007): S133-S140.
\bibitem[Garmire et al. (2003)]{Garmire2003} Garmire, Gordon P. Advanced CCD imaging spectrometer (ACIS) instrument on the Chandra X-ray Observatory. Proc. of SPIE 4851 (2003): 28-44.
\bibitem[Gunderson et al. (2000)]{Gunderson2000} Gunderson, Kurt S., Erik Wilkinson, and Jc Green. Calibrations and flight performance of the extreme ultraviolet opacity rocket. Proceedings of SPIE 4013 (2000): 421-425.
\bibitem[Hamilton et al. (1983)]{Hamilton1983} Hamilton, Ajs, Cl Sarazin, and Ra Chevalier. X-ray Line Emission from Supernova Remnants. I. Models for Adiabatic Remnants. The Astrophysical Journal Supplement Series 51 (1983): 115-148.
\bibitem[Katsuda et al. (2008)]{Katsuda2008} Katsuda, S, and H Tsunemi. XMM-Newton observations across the Cygnus Loop from northeastern rim to southwestern rim. Advances in Space Research 41, no. 3 (2008): 383-389.
\bibitem[Koyama et al. (1995)]{Koyama1995} Koyama, K, R Petre, Ev Gotthelf, U Hwang, M Matsuura, M. Ozaki, and S. S. Holt. Evidence for shock acceleration of high-energy electrons in the supernova remnant SN1006. Nature 378 (1995): 255-258.
\bibitem[Levenson et al. (1998)]{Levenson1998} Levenson, N.A., J.R. Graham, L.D. Keller, and M. J. Richter. Panoramic Views of the Cygnus Loop. The Astrophysical Journal Supplement Series 118 (October 1998): 541-561.
\bibitem[Levenson et al. (2005)]{Levenson2005} Levenson, N.A., and J.R. Graham. Environmental Impact on the Southeast Limb of the Cygnus Loop. The Astrophysical Journal 622 (March 2005): 366-376.
\bibitem[Liedahl et al. (1995)]{Liedahl1995} Liedahl, D A, A L Osterheld, and W H Goldstein. New Calculations of the Fe L-Shell X-ray Spectra in High-Temperature Plasmas. The Astrophysical Journal 438 (1995): 115-118.
\bibitem[Lisse et al. (1996)]{Lisse1996} Lisse, C.M., Dennerl, K., Englehauser, J., et al. Discovery of X-ray and Extreme Ultraviolet Emission from Comet C/Hyakutake 1996 B2, 1996, Science, 274, 205
\bibitem[McCammon et al. (2002)]{McCammon2002} McCammon, D, R Almy, E Apodaca, W Bergmann Tiest, W Cui, S Deiker, M Galeazzi, et al. A High Spectral Resolution Observation of the Soft X-ray Diffuse Background with Thermal Detectors. The Astrophysical Journal 576 (2002): 188-203.
\bibitem[McEntaffer et al. (2006)]{McEntaffer2006} McEntaffer, R L, W. Cash, A. Shipley, and E. Schindhelm. A sounding rocket payload for X-ray observations of the Cygnus Loop. Proc. SPIE 6266 (2006)
\bibitem[McEntaffer \& Cash (2008)]{McEntaffer2008} McEntaffer, R L, and W Cash. Soft x-ray spectroscopy of the cygnus loop supernova remnant. The Astrophysical Journal 680 (2008): 328-335.
\bibitem[McEntaffer et al. (2008)]{McEntaffer2008b} McEntaffer, R L, W Cash, and A Shipley. Off-plane reflection gratings for Constellation-X. Proceedings of SPIE 7011 (2008): 701107-1 - 701107-8.
\bibitem[McEntaffer et al. (2009)]{McEntaffer2009} McEntaffer, R L, N Murray, A Holland, C Lillie, S Casement, D Dailey, T Johnson, et al. Off-plane grating spectrometer for the International X-ray Observatory. Proceedings of SPIE 7437, no. 319 (2009): 74370H-74370H-13.
\bibitem[McEntaffer et al. (2010)]{McEntaffer2010} McEntaffer, R L, N. J. Murray, A. D. Holland, J. Tutt, S. J. Barber, R. Harriss, T. Schultz, et al. Developments of the off-plane x-ray grating spectrometer for IXO. Instrumentation 7732 (2010): 77321K-77321K-13. 
\bibitem[Oakley et al. (2009)]{Oakley2009} Oakley, P H H, W Cash, R L McEntaffer, A Shipley, and T Schultz. The EXOS sounding rocket payload. In Proceedings of SPIE, 7437:74370I, 2009.
\bibitem[Oakley et al. (2010)]{Oakley2010} Oakley, P H H, B Zeiger, M Kaiser, A Shipley, W Cash, R L McEntaffer, and T Schultz. Results from the Extended X-ray Off-plane Spectrometer (EXOS) sounding rocket payload. Instrumentation 7732, no. 2008 (2010): 77321R-77321R-8. 
\bibitem[Osterman et al. (2004)]{Osterman2004} Osterman, S, R L McEntaffer, W Cash, and A Shipley. Off-plane grating performance for Constellation-X. Proceedings of SPIE 5488 (2004): 302-312.
\bibitem[Reynolds \& Keohane (1999)]{Reynolds1999} Reynolds, S P, and J W Keohane. Maximum Energies of Shock-Accelerated Electrons in Young Shell Supernova Remnants. The Astrophysical Journal 525 (1999): 368-374.
\bibitem[Sanders et al. (1998)]{Sanders1998} Sanders, W., R. Edgar, D. Liedahl, and J. Morgenthaler. The Soft X-ray Background Spectrum from DXS. LNP 506 (1998): 8390.
\bibitem[Sanders et al. (2001)]{Sanders2001} Sanders, W T, Richard J Edgar, W L Kraushaar, D McCammon, and J P Morgenthaler. Spectra of the 1/4 keV X-ray Diffuse Background from the Diffuse X-ray Spectrometer Experiment, The Astrophysical Journal, 554 (2001): 694-709.
\bibitem[Snowden et al. (2004)]{Snowden2004} Snowden, S. L., M. R. Collier, and K. D. Kuntz. XMM-Newton Observation of Solar Wind Charge Exchange Emission. The Astrophysical Journal 610, no. 2 (August 2004): 1182-1190.
\bibitem[Snowden et al. (1995)]{Snowden1995} Snowden, S L, M J Freyberg, P P Plucinsky, J. Schmitt, J. Tr\"umper, W. Voges, R J Edgar, D. McCammon, and W T Sanders. First Maps of the Soft X-ray Diffuse Background from the ROSAT XRT/PSPC All-Sky Survey. The Astrophysical Journal 454 (1995): 643. 
\bibitem[Snowden et al. (1997)]{Snowden1997} Snowden, S L, R Egger, M J Freyberg, D Mccammon, P P Plucinsky, W T Sanders, J H M M Schmitt, J Tru, and W Voges. ROSAT Survey Diffuse X-ray Background Maps. II. The Astrophysical Journal 485 (1997): 125-135.
\bibitem[Tamagawa (2006)]{Tamagawa2006} Tamagawa, T. Fine-pitch and thick-foil gas electron multipliers for cosmic x-ray polarimeters. Proceedings of SPIE 6266, no. 2006 (2006): 62663W-62663W-10.
\bibitem[Tamagawa et al. (2008)]{Tamagawa2008} Tamagawa, T, A Hayato, K Abe, S Iwamoto, S Nakamura, A Harayama, T Iwahashi, K Makishima, H Hamagaki, and Y Yamaguchi. Gain properties of gas electron multipliers (GEMs) for space applications. Proc. of SPIE 7011 (2008): 1-8.
\bibitem[Tamagawa et al. (2009)]{Tamagawa2009}Tamagawa, T, A Hayato, F Asami, K Abe, S Iwamoto, S Nakamura, A Harayama, et al. Development of thick-foil and fine-pitch GEMs with a laser etching technique. Nuclear Instruments and Methods in Physics Research 608 (2009): 390-396.
\bibitem[Tsunemi et al. (2009)]{Tsunemi2009} Tsunemi, H, M Kimura, H Uchida, K Mori, and S Katsuda. Another Abundance Inhomogeneity in the South East Limb of the Cygnus Loop. Publications of the Astronomical Society of Japan 61 (2009): S147-S153.
\bibitem[Werner (1977)]{Werner1977} Werner, W. X-ray efficiencies of blazed gratings in extreme off-plane mountings. Applied optics 16, no. 8 (August 1977): 2078-80.
\end{thebibliography}
\end{document}